\documentclass{emulateapj}
\usepackage{graphicx}

\shorttitle{Gravitational Lens Modeling with GAs and PSOs}
\shortauthors{Rogers \& Fiege}

\begin{document}
\textheight 9.0in

\title{Gravitational Lens Modeling with Genetic Algorithms and Particle Swarm Optimizers}
\author{Adam Rogers \& Jason D. Fiege}
\affil{Department of Physics and Astronomy, The University of
Manitoba, Winnipeg, Manitoba, Canada R3T-2N2}
\email{rogers@physics.umanitoba.ca}
\slugcomment{To appear in ApJ, 727:80, Feb.1, 2011. Submitted 2010 July 9; accepted 2010 Nov 19}

\begin{abstract}
Strong gravitational lensing of an extended object is described by a mapping from source to image coordinates that is nonlinear and cannot generally be inverted analytically. Determining the structure of the source intensity distribution also requires a description of the blurring effect due to a point spread function.  This initial study uses an iterative gravitational lens modeling scheme based on the semilinear method to determine the linear parameters (source intensity profile) of a strongly lensed system.  Our `matrix-free' approach avoids construction of the lens and blurring operators while retaining the least squares formulation of the problem. The parameters of an analytical lens model are found through nonlinear optimization by an advanced genetic algorithm (GA) and particle swarm optimizer (PSO).  These global optimization routines are designed to explore the parameter space thoroughly, mapping model degeneracies in detail.  We develop a novel method that determines the L-curve for each solution automatically, which represents the trade-off between the image $\chi^2$ and regularization effects, and allows an estimate of the optimally regularized solution for each lens parameter set.  In the final step of the optimization procedure, the lens model with the lowest $\chi^2$ is used while the global optimizer solves for the source intensity distribution directly.  This allows us to accurately determine the number of degrees of freedom in the problem to facilitate comparison between lens models and enforce positivity on the source profile.  In practice we find that the GA conducts a more thorough search of the parameter space than the PSO.
\end{abstract}

\section{Introduction}

Strong gravitational lens effects produce multiple distorted images of a background object and also provide magnification of lensed sources.  Magnification may reveal unresolved features in lensed sources and provides a useful tool for studying cosmologically distant objects.  Furthermore, gravitational lensing provides a unique method to determine the mass distribution of lensing objects, which can be most accurately modeled when the lens potential is probed in parallel at many points.  Therefore accurate lens inversion methods for extended sources are important because they provide a large number of constraints on the lens mass distribution.

Models of both the intensity profile of the source and the lens mass distribution are required to model a strong gravitational lens system.  Analytical models of the source are sometimes used because they are typically described by a small number of parameters, and can ensure smoothness and positivity when used to model the source intensity distribution. However, the correct parameterization is not always clear for such models, and the choice of a specific parametric model biases the lens and source solutions.  Authors have attempted to partially overcome this drawback by using complex but flexible parametric models specified by large parameter sets.  The most extreme example is \citet{Tyson}, who used an elaborate source model with more than $200$ parameters to fit the gravitational lens $CL0024 +1654$.  In such cases, it may be simpler to use pixelized source models, which treat all pixels on the source independently.

The difficulty with pixelized source models is that they require many more free parameters than even the most complex analytical models.  The optimization of extended sources via pixelized intensity distributions is simplified using the versatile semilinear method developed by \citet{WD03} and later expanded upon by a number of authors, including \citet{TK04}.  The semilinear method uses a pixelized source, and also incorporates the blurring due to the point-spread function (PSF) of the instrument used to obtain the data.  Additive noise in the observational data is also taken into account by the semilinear method.

In this paper we detail Mirage, a gravitational lens modeling code written in MATLAB and C.  The present version of Mirage is designed to optimize the parameters of analytical lens models and pixelized sources, but work is underway to extend the code to handle non-parametric lens models as well.  A modified version of the semilinear method forms the backbone of our lens modeling program. We use sophisticated global optimization methods to fit the lens parameters, and the semilinear method to determine the corresponding source light profile that best matches the data.  As a final step, we employ the method of \citet{BLGA} to enforce the positivity of the source while keeping the nonlinear lens parameters constant.  This affords a method of comparison between distinct lens density models because the number of degrees of freedom is well-defined and fixed \citep{BL06}.  The global optimizers studied in this paper consist of a sophisticated genetic algorithm (GA), called Ferret \citep{fiege04}, and an enhanced particle swarm optimizer (PSO), Locust, which are both components of the Qubist Global Optimization Toolbox by \citet{Fiege}.  This paper discusses a robust method for gravitational lens reconstructions, highlights the benefits of both types of optimization routines, and compares their performance.

In Section \ref{sec:lensing} we will review the gravitational lens inverse problem, the semilinear method and our new matrix-free approach to lens modeling. In Section \ref{sec:fullopt} we discuss the details of the GA and PSO, as well as a variety of simulated data tests. Section \ref{sec:results} presents our results using these methods, and our conclusions are summarized in Section \ref{sec:conclusions}.

\section{The Gravitational Lens Problem}
\label{sec:lensing}

In this section, we describe the background of the gravitational lens problem. Section \ref{sec:SL} reviews the semilinear method, and Section \ref{sec:matfree} details our matrix-free method. A small scale test of our algorithm is discussed in Section \ref{sec:SmallTest} and the implicit regularizing properties of iterative methods are described in Section \ref{sec:implicitReg}. Determination of the effective number of degrees of freedom is discussed in Section \ref{sec:MCDOF}  and the L-Curve criteria in Section \ref{sec:Lcurve}. Finally, the results of our algorithm applied to a large-scale test are shown in Section \ref{sec:LargeScaleTest}.

We invoke the standard thin lens geometry \citep{refsdal}, tiling the source plane with coordinates {\boldmath $\beta$}=($\xi$,$\eta$) and the observed image plane with coordinate system {\boldmath$\theta$}=($x$,$y$)  \citep{schneider}. The thin lens equation maps light rays from the image plane to the source plane, such that
\begin{equation}
\mbox{\boldmath$\beta$} = \mbox{\boldmath$\theta$} -
\mbox{\boldmath$\alpha$}(\mbox{\boldmath$\theta$}).
\label{eq:thinlens}
\end{equation}
Equation (\ref{eq:thinlens}) is nonlinear because the deflection angle depends on the image coordinates, and multiple solutions to the lens equation may exist when it is solved \citep{SEF}.  This multi-valued property makes the lens mapping non-invertible in general, and the lens equation can be solved analytically for only a handful of simple lens models.

It is straightforward to find the gravitationally lensed image of a background source in the absence of blurring, given a model of the source and a lens mass distribution.  The deflection angle {\boldmath $\alpha$($\theta$)} determines the position of image pixels back-traced to the source plane, and the brightness of each image pixel is determined by conservation of surface brightness:
\begin{equation}
S(\mbox{\boldmath$\theta$})=\Sigma(\mbox{\boldmath$\beta$}(\mbox{\boldmath$\theta$})) \label{consSurfB},
\end{equation}
where S({\boldmath$\theta$}) represents the intensity at point $(x,y)$ on the
image plane and $\Sigma${\boldmath$(\beta$}({\boldmath$\theta$})) is the corresponding back-traced
intensity at position $(\xi(${\boldmath$\theta$}), $\eta(${\boldmath$\theta$})) on the source plane \citep{KS}.  In principle, it is also possible to use Equation (\ref{consSurfB}) as a simple method to calculate the intensity distribution of the source from an image.  However, methods based on the conservation of surface brightness are complicated by the fact that many back-traced image pixels may land within any given source pixel due to the multiple imaging property of the lens mapping.  When this occurs, the source pixel is assigned the mean intensity of the back-traced image points.  This approach to lens modeling is called the Digital Source Reconstruction (DiSoR) method \citep{KS}, and derives the structure of a lensed source based on the observed pixel intensities given a lens model.  The advantage of this approach is that the conservation of surface brightness (Equation (\ref{consSurfB})) avoids solving the lens equation directly, since this would in general involve finding the solutions to a complicated nonlinear equation at the position of each source pixel \citep{SK}.

Despite their tempting simplicity, lens inversion schemes based on the DiSoR method cannot be used for most applications because they fail in the presence of significant distortion due to instrumental and atmospheric blurring described by a PSF.  Readout and background noise further complicate the application of the conservation of surface brightness by corrupting the pixel intensities present in the data.  These factors make it impossible to use the DiSoR approach to obtain the exact inversion of the lens system through ray-tracing, because multiply imaged pixels may no longer have identical intensities, which invalidates the fundamental assumption of the method. When significant blurring or noise is present, a forward modeling approach is needed such that a source model is lensed using Equation (\ref{consSurfB}) and convolved with the PSF for comparison with the data.  The semilinear method is based on forward modeling and reduces the lens inversion to a least-squares type of problem.

\subsection{The Semilinear Method}
\label{sec:SL}

The semilinear method provides a way to solve for optimal source intensities by the direct inversion of a lens matrix, for a given set of lens parameters.  However, the search for optimal lens parameters is nonlinear in general, and requires more sophisticated nonlinear optimization methods, such as those discussed in Section \ref{sec:global}.  Fast execution of the matrix inversion part of the problem is crucial, because a linear system of equations must be solved for each set of lens parameters tested by the nonlinear optimizer during the search for optimal lens models. Many sets of lens parameters must be evaluated to search the parameter space thoroughly enough to determine the globally optimal solution.

Following \citet{WD03}, we label the image pixels $j=1..J$ and treat the pixels in the source as independent free parameters $i=1..I$. Given a set of lens parameters, the image of each source pixel is formed by ray-tracing assuming unit brightness $s_i=1$, and convolved with the PSF.  This transformation is encoded in a matrix {\boldmath$f$}={\boldmath$BL$}.  We assume linear blurring described by the blurring matrix {\boldmath$B$} which accounts for the PSF.  The matrix {\boldmath$L$} performs ray-tracing via the lens equation (\ref{eq:thinlens}).  The problem is then reduced to finding a set of source pixel scaling factors $s_i$ that minimize the reduced $\chi^2$ statistic between the model image and the observed data.  Using the set of source pixel intensities, the lensed image of a source is found easily:
\begin{equation}
b_j=\sum_i s_i f_{ij},
\label{mapping}
\end{equation}
where $f_{ij}$ are the elements of the matrix {\boldmath$f$}.  The $\chi^2$ statistic between the lensed image and the data is:
\begin{equation}
\chi^2=\sum_j \frac{\left( \sum_{i} s_{i}f_{ij}-d_j
\right)^2}{\sigma_{j}^2} \label{chi2}
\end{equation}
where $d_j$ are the observed intensity in each image pixel, and $\sigma_j$ is the standard deviation error associated with pixel $j$.  After differentiating this equation with respect to the source pixel intensities $s_i$, we define $\mbox{\boldmath$F$}=f_{ij}/\sigma_j$, $\mbox{\boldmath$\hat{d}$}=d_j/ \sigma_j$ and we obtain the relation
\begin{equation}
\mbox{\boldmath$F$}^{T}\mbox{\boldmath$F$}\mbox{\boldmath$s$}=\mbox{\boldmath$F$}^{T}\mbox{\boldmath$\hat{d}$},
\label{eqLS}
\end{equation}
from which it follows that the source pixel scalings can be determined by linear inversion:
\begin{equation}
\mbox{\boldmath$s$}=\mbox{\boldmath$M$}^{-1}\mbox{\boldmath$\hat{d}'$}
\end{equation}
where {\boldmath$\hat{d}'$}$=$\mbox{\boldmath$F$}$^{T}$\mbox{\boldmath$\hat{d}$} and {\boldmath$M$}={\boldmath$F$}$^{T}${\boldmath$F$}.  This inversion determines the optimal set of source pixel scalings necessary to reproduce the observed data for a given lens model.  Further details of this derivation can be found in \citet{WD03} and \citet{TK04}.  In the standard semilinear method, the system matrix {\boldmath$M$} is very large, where the linear size of the matrix scales as the number of source pixels used in the inversion.  The matrix is very sparse when the PSF is narrow, but a greater fraction of matrix elements are non-zero for increasingly broad PSFs.  Mirage uses sparse matrix methods to minimize memory usage.

\citet{WD03} originally presented the semilinear method as a ray-shooting algorithm that performs nearest neighbor interpolation. \citet{TK04} modified the lens matrix to accommodate bilinear interpolation of the source. This consists of using the four source pixels surrounding a back-traced image pixel with appropriate weighting to assign a brightness value to each image pixel.  Mirage currently implements nearest neighbor, bilinear, and bicubic source plane interpolation.  Higher order interpolation schemes are possible, but they are computationally more expensive and result in a lens mapping matrix that is less sparse.  It is also possible to use more elaborate schemes to grid the source plane, including the Delaunay tesselation scheme used by \citet{VK09}.  We restrict the source models to rectilinear grids in this paper, but plan to explore other such options.

In practice, regularization is necessary to stabilize the matrix inversion due to the presence of noise in the data \citep{TK04}.
This regularization term makes the system matrix {\boldmath$M$} more diagonally dominant and hence better conditioned, which has the effect of increasing the smoothness of the source light distribution.  In general, we add a regularization matrix to the system matrix, to give
\begin{equation}
\mbox{\boldmath$M'$}=\mbox{\boldmath$M$}+\lambda\mbox{\boldmath$H$}^T\mbox{\boldmath$H$} \label{TKReg},
\label{L}
\end{equation}
where $\lambda$ is an adjustable regularization parameter. It is then possible to control the smoothness of the derived solution by adjusting the regularization parameter, with the unregularized case recovered as $\lambda \rightarrow 0$.  Zeroth order regularization has \mbox{\boldmath$H$}=\mbox{\boldmath$I$}, which suppresses noise in the inversion by preferring sources with less total intensity \citep{WD03}.  It is also possible to introduce more complicated forms of regularization, typically based on finite difference representations of two-dimensional derivative operators.

It has been shown that different regularizing terms produce qualitatively similar results \citep{TK04}, and the behavior of a host of linear regularization schemes has been studied in great detail by \citet{suyu06} in the framework of Bayesian analysis.  An important drawback of regularization is that it introduces dependencies between source plane pixels, which makes it difficult to characterize the effective number of degrees of freedom required to compute the reduced $\chi^2$ ($\chi^2_r$).  Therefore, direct comparison of different models is more difficult in regularized schemes than without regularization.  \citet{DW05} use an adaptive mesh in the source plane to overcome the problem of calculating the number of degrees of freedom in the problem.

An important advantage of the semilinear method is that errors of the source intensity parameters can be easily determined from the lensing matrix, as seen from the relation
\begin{equation}
M_{ik}=\frac{1}{2}\frac{\partial^2 (\chi^2)}{\partial s_i \partial s_k}.
\end{equation}
This equation expresses the lensing matrix as half of the Hessian matrix of the reduced image $\chi^2$ statistic. \citet{WD03} use this relationship to find the covariance matrix {\boldmath$C$}={\boldmath$M$}$^{-1}$, thus determining the source plane errors automatically during application of the semilinear method.  When regularization is used, the covariance matrix cannot be found in this way, but \citet{WD03} proposed a Monte Carlo method as an alternative method to estimate errors.

Despite its conceptual elegance, there are significant practical limitations and drawbacks to the semilinear method.  The number of non-zero matrix elements of {\boldmath$M$} scales linearly with the number of pixels in the source and depends on the source interpolation method used.  Direct inversion quickly becomes impractical for very large images, or those with large PSFs. This sparsity requirement can be fulfilled by representing the PSF by a simpler function, for example a Gaussian, and setting small values to zero. This thresholding helps to control the potentially poor conditioning of the blurring matrix.  However, realistic PSFs may contain significant low-level structure, and fitting a simple analytical function to it may not be desirable in such cases.  The semilinear scheme is therefore most practical when the image is small and the PSF is narrow, as in typical optical data.

Another problem with the semilinear method is that it does not enforce the positivity of source pixel intensities.  Optimal source solutions derived using this algorithm may contain negative pixel intensities, due to noise in the data, since bounds cannot be enforced in the matrix inversion step.  Moreover, there is no form of linear regularization that is guaranteed to prevent this behavior.  We note, however, that it is possible to enforce positivity in other lens modeling schemes, such as the maximum entropy method (MEM), explored by \citet{Wallington}.

In summary, semilinear inversion provides a convenient method for modeling strongly lensed extended sources because it states the gravitational lens modeling problem using a least-squares approach that is solved by direct matrix inversion.  The inversion step guarantees that the globally optimal solution is found for unbounded source pixel values.  However, the method is computationally expensive for large images and PSFs.  In such cases even building the transformation matrix {\boldmath$f$}, incorporating both lensing and blurring effects, is an expensive computation and the inversion step may be time-consuming or impractical due to the poor sparsity and size of the matrix.  We show in Section \ref{sec:matfree} that it is possible to derive a `matrix-free' formulation that avoids the explicit construction of the matrix and dramatically improves the efficiency of solving for the linear parameters by employing local optimization methods to solve the least squares problem.  We also compare results from this technique with the semilinear inversion method and show that both methods produce solutions of similar quality. A final refinement step ensures positivity of the source pixels, thus rectifying a limitation of the standard semilinear method.

\subsection{Matrix-free modeling of lensed images}
\label{sec:matfree}
The main goals of Mirage are generality, flexibility and sufficient efficiency to allow thorough exploration of lens model parameter space by global optimization techniques, which typically require the evaluation of $10^4-10^5$ lens models.  We therefore require a fast code to solve the linear part of the problem, which is able to function with arbitrarily complicated PSFs and data of high resolution. Mirage implements the semilinear method, using direct matrix inversion, but also extends this method by using faster and less memory intensive local iterative optimization routines that avoid the need for explicit construction of the lens and blurring operators. The iterative methods in Mirage are not intended as a replacement for the semilinear method, which is the preferred approach when it is computationally practical.  However, memory requirements and long run times for the global lens parameter search may practically restrict the semilinear method to source intensity distributions and PSFs that can be built on a small mesh to limit the matrix size and maintain its sparsity.  Our technique is intended to augment the least-squares approach of the semilinear method by providing a complement of algorithms capable of modeling large lens images quickly, even if the PSF is also large.  We avoid the direct inversion of large matrices, but maintain the least-squares formulation, allowing the use of any linear optimization algorithm suited to the solution of large-scale problems.  Since this paper focuses on solving the full nonlinear lens modeling problem, we largely make use of matrix-free methods for the remainder of this work except for comparisons with the direct semilinear method.

Given the parameters of a lens model and assuming a source intensity distribution, matrix multiplication with {\boldmath$L$} results in the unblurred lensed image. This image can also be found by the conservation of surface brightness, as given by Equation (\ref{consSurfB}). By storing the positions of the back-traced image pixels on the source plane, we perform an interpolation on the source plane directly, which allows us to find the unblurred lensed image without the need for an explicit representation of the lens matrix. Similarly, a separate algorithm in Mirage has an effect equivalent to multiplying by the transpose of the lens matrix, which works by carefully keeping track of the positions of back-traced image pixels.

The lens mapping magnifies portions of the image plane by differing amounts such that square image pixels mapped back to the source plane may no longer remain square. This effect is especially pronounced when image pixels are traced back to the source plane near caustic curves. In general, the distortion in shape of the back-traced image pixels may cause portions of the back-traced pixels to lie within separate source pixels. To account for this effect, we split each image pixel into $N_{p}$ subpixels, and trace each of these subpixels to the source plane independently. By interpolating each of these subpixels on the source mesh, we can then average over their intensities in the image plane to find a better estimate of the lensed intensity profile. $N_p$ can be any size, but the execution time of the code increases as we include finer subpixel resolution. This approach improves the accuracy of the transformation from the lens to source plane.  In addition, we find that it improves the smoothness of the $\chi^2$ surface, which helps with the global search for optimal lens parameters. Image pixels that do not map to the source plane are not included in the local optimization and are assigned no intensity.

To successfully model a realistic blurred observation, we incorporate the blurring effect of the PSF, which is usually described by the blurring matrix {\boldmath$B$}. This matrix is of size $N_i^2 \times N_i^2$ where $N_i$ is the number of image pixels which map back to the source plane. Since the lens matrix {\boldmath$L$} is sparse, avoiding its construction does not directly increase the speed of the code, but it allows us to sidestep the explicit construction of the blurring matrix. Since PSFs may describe a significant blurring effect, the product of the lens and blurring matrices {\boldmath$f$} may have a large number of non-zero entries, decreasing the sparsity of the system.  To avoid building the blurring matrix, we convolve with the PSF in Fourier space, which is computationally inexpensive, even for large images.  Convolution without a blurring matrix is common in the image processing community and was used by \citet{NagyRestoreTools} to solve least-squares problems in the context of the standard image deconvolution problem. By direct extension, this `matrix-free' lensing method allows us to solve the least-squares problem described by \citet{WD03} without the need for explicit representation of the matrices involved using local optimization. This approach not only provides a substantial increase in speed, but also allows for the use of large data sets with complicated PSFs.  In principle the matrix-free approach can be extended to spatially variant PSFs using the techniques described by \citet{NagyRestoreTools}.

The local linear optimization algorithms considered in this paper require an initial guess of the solution.  We begin with a blank source prior, and each successive iteration adds increasingly higher spatial frequency detail to this initial guess.  Local optimizers such as the conjugate gradient method for least squares problems (CGLS; \citep{bjorck}) and steepest descent (SD) require one matrix multiplication and one matrix transpose multiplication per iteration, so that the least squares problem can be solved without explicit representation of the lens matrix.  We show that such a procedure converges in practice to a source model that is very close to the solution found through matrix inversion, given a sufficient number of iterations.  Moreover, an explicit regularization term is not required in general since iterative optimization techniques have been shown to have an automatic regularizing effect on the problem \citep{vogel}, allowing Equation (\ref{eqLS}) to be minimized directly.

Iterative schemes have been used previously in the strong lensing literature in application to nonlinear regularization, for example by \citet{Wallington} in the lensMEM method. A similar approach is used in the LENSVIEW code by \citet{lensView}, which utilizes the MEM discussed by \citet{SBMEM}. The semilinear method is restricted to linear regularization terms of the type detailed in \citet{suyu06}. Nonlinear regularization like the MEM can also be used with Mirage, although these techniques require more complicated nonlinear optimization schemes to solve for the source intensity distribution.

\subsection{A Small-scale Test}
\label{sec:SmallTest}
\begin{figure}
\scriptsize \epsscale{1} \plotone{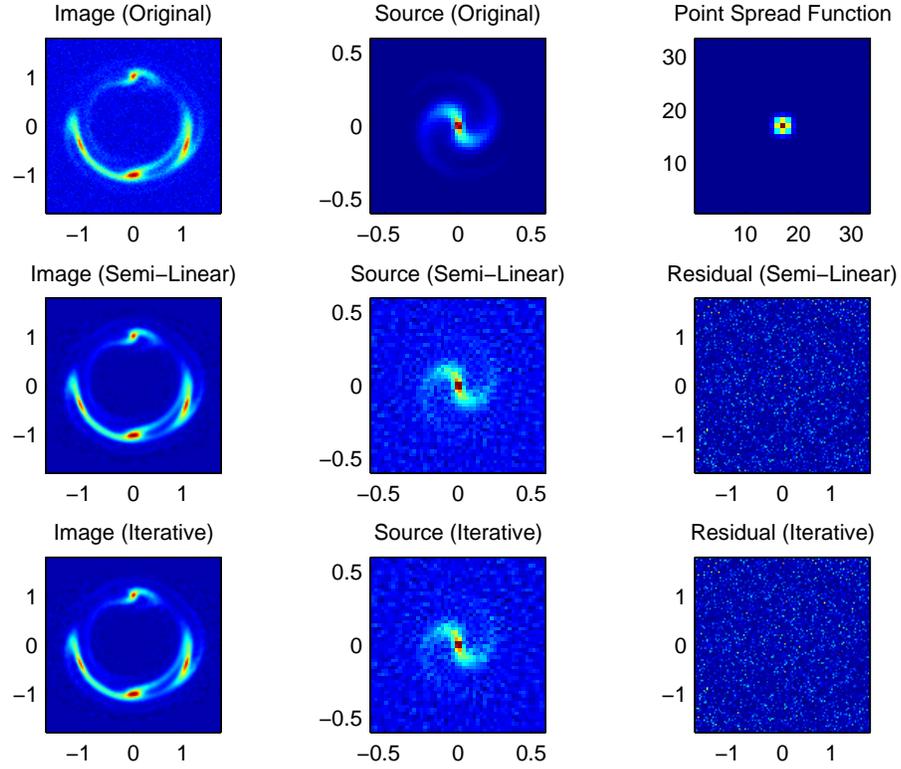} \caption{Top row, from left: artificial data, source intensity distribution and Gaussian PSF used to generate the observation. The source was built on a $50 \times 50$ grid, and the lensed image is defined on a $120 \times 120$ grid. Middle row, from left: model observation of resulting source intensity reconstruction, source intensity profile as found by the semilinear method, and the resulting image residuals. Zeroth order regularization with a regularization constant $\lambda=2.5 \times 10^{-3}$ was used to reconstruct the source. Bottom row, from left: resulting model image, model source and image residuals as determined by the CGLS algorithm after $40$ iterations. Note the similarity between the semilinear and iterative solutions with respect to the derived source. Although both of these models contain back-traced noise, the real features of the source are reproduced and clearly visible in the reconstructions.} \label{compareImg}
\end{figure}

As a small scale test problem, we generated a $120 \times 120$ pixel image of an analytical source intensity profile. The image pixel size used in this test is $0.03$ arcsec. The test source is defined by a two-armed spiral test function, given by \citet{Bonnet}:
\begin{equation}
S(r,\omega)=\frac{S_0}{r_c^2+r^2}exp{\left[-2~\sin^2\left(\omega-\omega_0-\tau r^2\right)\right]},
\label{bonner}
\end{equation}
where $S_0$ is the maximum brightness in arbitrary units and core radius $r_c$.  The tightness of the arms about the central bulge is controlled by $\tau$, and $\omega_0$ controls the orientation of the spiral, in standard polar coordinates $\left(r,\omega\right)$. The lensed image of this function can be complicated, since the function contains significant structure, and therefore provides a good test of the level of detail that we are able to recover using our lens inversion algorithm. A test image is generated using the approach detailed above, with each image pixel composed of a $10 \times 10$ grid of subpixels.  The high subpixel resolution mimics the smooth structure of a natural image. We blur the resulting image by convolving with a Gaussian PSF, with a FWHM of $5$ pixels on a $30 \times 30$ grid, and add Gaussian noise to construct our final test image, as shown in Figure \ref{compareImg}.  Our simulated data set has a signal-to-noise ratio $S/N=8$, where we define the S/N as the maximum image intensity divided by the standard deviation of the additive Gaussian noise.  The lens used to produce this image is the six-parameter singular isothermal ellipse (SIE; \citet{keeton}) which has Cartesian deflection angles given by
\begin{equation}
\alpha_x=\frac{bq}{\sqrt{1-q^2}}\tan^{-1}\left( \frac{x\sqrt{1-q^2}}{\psi+s} \right)
\label{SIEX}
\end{equation}

\begin{equation}
\alpha_y=\frac{bq}{\sqrt{1-q^2}}\tanh^{-1}\left( \frac{y\sqrt{1-q^2}}{\psi+q^2s} \right),
\label{SIEY}
\end{equation}
where $\psi^2=q^2(s^2+x^2)+y^2$ and $q=\sqrt{ (1-\epsilon)/(1+\epsilon)}$, and $b$ is the corresponding Einstein radius in the limit of a spherical model with $q=1$.  The parameter $b$ is related to the velocity dispersion $\sigma_v$ by
\begin{equation}
b=4\pi\left(\frac{\sigma_v}{c}\right)^2\frac{D_{ls}}{D_{os}}
\label{einsteinRadius}
\end{equation}
where $D_{ls}$ and $D_{os}$ are the angular distances between lens and source and observer and source respectively, and $c$ is the speed of light.  The actual parameters used to construct Figure \ref{compareImg} are as follows: The velocity dispersion is $\sigma_v=260$ km s$^{-1}$ resulting in $b=1.35$ arcsec, ellipticity $\epsilon=0.4$, lens center $(x,y)=(0,0.12)$, orientation angle $\theta_L=\pi/2$.  We keep the core size fixed at $s=0$.  In addition to these six parameters, we assume that the redshift of the lens and source are $z_d = 0.3$ and $z_s = 1.0$ respectively.  For convenience we measure angular distances with respect to the ``flat'' Friedmann metric with $k=0$.

\begin{figure}
\scriptsize \epsscale{1} \plotone{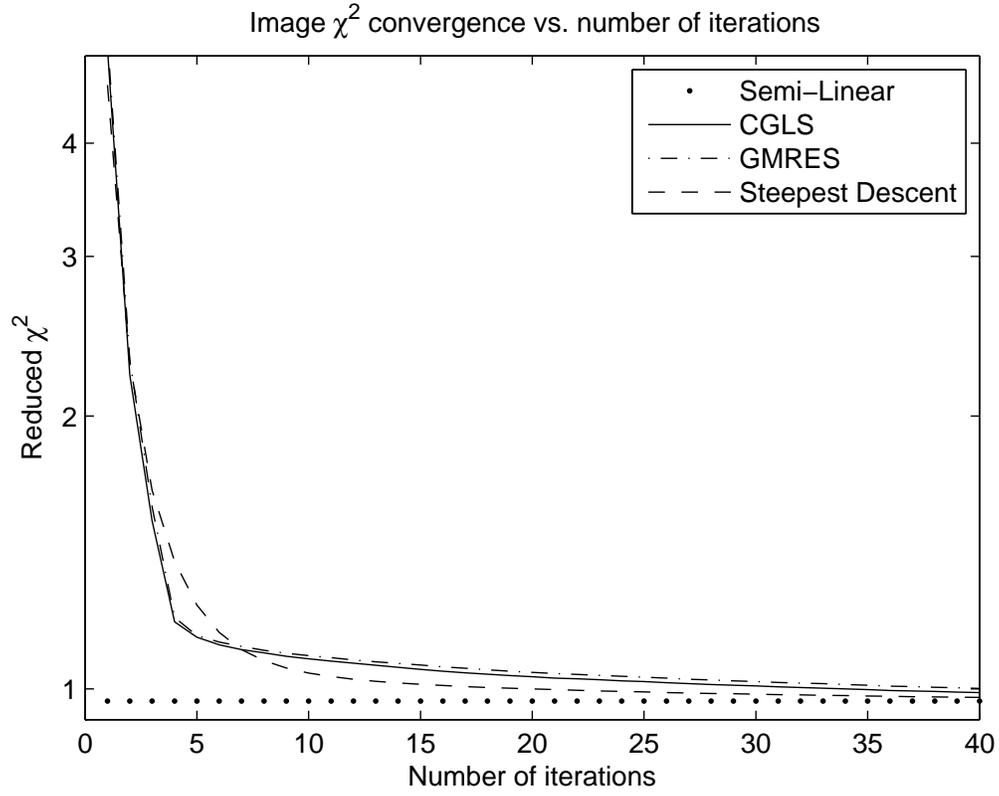} \caption{Convergence properties of several local optimization routines. The CGLS and LSQR \citep{bjorck} algorithms exhibit identical behavior, and the performance of the GMRES \citep{saad} algorithm is similar. The steepest descent algorithm converges more slowly but attains a slightly lower image $\chi^2$ value. The difference between the local optimization routines and the semilinear method is emphasized on this plot due to the logarithmic scale.  The semilinear method result was obtained using a zeroth order regularization constant $\lambda=2.5 \times 10^{-3}$, and iterative algorithms were terminated after $40$ iterations.}
\label{convergeImg}
\end{figure}

We model the data using the semilinear method and matrix free methods with subpixel grids of size $2 \times 2$. The sub-pixel resolution used for modeling is lower than that used to produce the data, which makes the test more realistic.  The source plane is defined on a $40 \times 40$ grid, with source pixels of size $0.024$ arcsec.  The original simulated-data image is shown in the first row of Figure \ref{compareImg}, the semilinear reconstruction on the second row, and a reconstruction using the CGLS algorithm on the third row. CGLS was chosen as the linear optimizer because of its speed and popularity as a local optimization scheme, but in practice all of the optimizers included in the Mirage package produce similar results. All of the local optimization algorithms tested are able to recover the details of the original source function well. Figure \ref{convergeImg} shows that the rate of convergence varies between optimization algorithms, but they all settle down to the minimum reduced image $\chi^2_r$ values to within $5 \%$ by generation $40$. The semilinear method is displayed on this plot simply as a constant $\chi^2_r$ because it is a direct method.  Note that all of the source reconstructions show noise back-traced from the image, which is unavoidable using pixel mapping techniques on data containing noise. All local optimization algorithms converge in approximately $2$ s, while  the semilinear method required approximately $16$ s.  This test was conducted on a $2.4$ GHz dual-core Intel machine with $3$ GB of memory.  Memory usage was monitored and did not exceed the hardware memory limit at any time.

\begin{figure}
\scriptsize \epsscale{1} \plotone{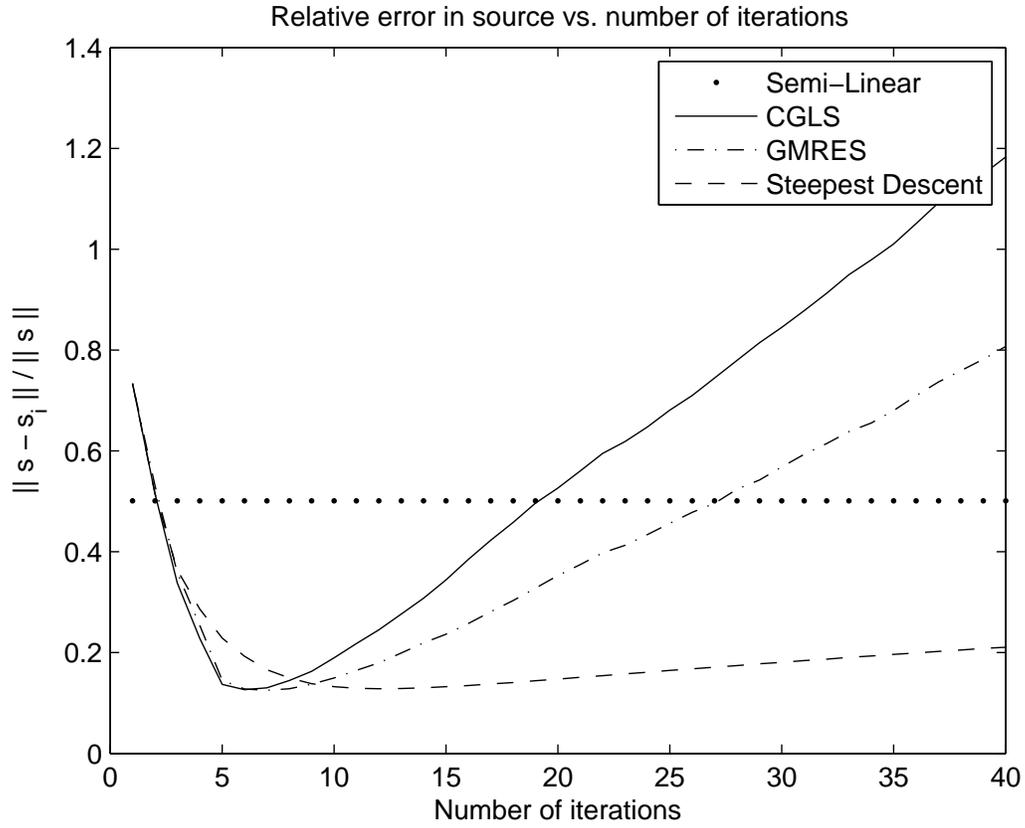} \caption{Convergence properties of the source intensity distribution. This figure plots the relative error between a given iterative method and the true source intensity distribution. All of the iterative optimization algorithms display semi-convergence behavior. The semilinear method result was obtained using a zeroth-order regularization constant $\lambda=2.5 \times 10^{-3}$, and iterative algorithms were terminated after $40$ iterations. The relative error of the solution found by the SD algorithm increases more slowly past convergence than the error for any of the other iterative schemes shown here.} \label{convergeSrc}
\end{figure}

In general, the lensed model image becomes increasingly well matched to the data the longer an iterative optimizer runs, but the usefulness of the solutions eventually starts to degrade as the algorithm begins fitting to the noise in the data.  Therefore, the corresponding noise level in the source reconstruction rises as iterations continue, which we can quantify for this test problem because we know the true solution in the absence of noise as shown in Figure \ref{convergeSrc}.  In effect, the number of iterations of the local optimizer acts as a regularization parameter \citep{fleming2}.  Thus, it is possible to introduce regularization by carefully controlling the number of iterations during local optimization.  In general, implicit regularization is present whenever these local optimizers are used in the context of deblurring problems, which implies that suitable stopping criteria must be established to find the optimally regularized solution \citep{hansen}. It is noteworthy that the semilinear method suffers from a related problem since the regularization constant is a free parameter, and therefore the associated ambiguity is equivalent to the problem of choosing a stopping criteria in iterative methods.  Techniques have been developed to deal with this problem using Bayesian analysis for the semilinear method, as discussed by \citet{BL06} and \citet{suyu06}.  For iterative methods, the issue of a stopping criteria is a non-trivial problem that has no unique solution for local optimization, although many methods exist to deal with this problem, such as Generalized Cross Validation \citep{wahba} and the L-curve criterion \citep{engel}.  We discuss a novel approach in Section \ref{sec:Lcurve} that uses the L-curve analysis to estimate the optimal regularization parameter (stopping condition) in conjunction with global parameter search methods.

\subsection{Iterative Optimization as Implicit Regularization}
\label{sec:implicitReg}
To see how iterative schemes produce implicit regularization, consider a system {\boldmath$b$}={\boldmath$Bx$}+{\boldmath$n$}, where {\boldmath$n$} describes the noise added to the true image. Suppose that the blurring matrix {\boldmath$B$} is ill-posed \citep{hansen97}, and the ``true'' solution is the unblurred image, represented as a vector {\boldmath$x$}.  Given the blurring matrix and the noisy data {\boldmath$b$}, we can formally write an approximate solution to the inverse problem as {\boldmath$x$}$=${\boldmath$B$}$^{-1}${\boldmath$b$}. However, this proves to be difficult in practice because of the poor conditioning of {\boldmath$B$} and the noise contained in the data. The resulting solution is the sum of two terms, {\boldmath$x$}$_n=${\boldmath$x$}$+${\boldmath$B$}$^{-1}${\boldmath$n$}. The second term can dominate the first in this expression, which results in poor recovery of the true solution, {\boldmath$x$}. To overcome this problem, regularization schemes seek a solution to the system
\begin{equation}
\mbox{\boldmath$x$}_{\lambda} = argmin \left( || \mbox{\boldmath$b$} - \mbox{\boldmath$B$}\mbox{\boldmath$x$} ||^2 + \lambda || \mbox{\boldmath$H$} \left( \mbox{\boldmath$x$}-\mbox{\boldmath$x$}_0 \right) ||^2 \right)
\label{linsys}
\end{equation}
where \mbox{\boldmath$H$} is the regularization matrix, \mbox{\boldmath$B$} is the system matrix and $b$ is a vector of data to be fit. The regularized solution is \mbox{\boldmath$x$}$_{\lambda}$, and the default solution is \mbox{\boldmath$x$}$_0$, which is found when $\lambda \rightarrow \infty$. By requiring the derivative of Equation (\ref{linsys}) to vanish we derive the following expression
\begin{equation}
\left( \mbox{\boldmath$B$}^T\mbox{\boldmath$B$} + \lambda \mbox{\boldmath$H$}^T\mbox{\boldmath$H$}\right) \mbox{\boldmath$x$}=\lambda\mbox{\boldmath$H$}^T\mbox{\boldmath$H$}\mbox{\boldmath$x$}_0+\mbox{\boldmath$B$}^T\mbox{\boldmath$b$}
\end{equation}
provided that the regularization is linear in nature. Note that the first term on the right depends on the default solution \mbox{\boldmath$x$}$_0$, which represents a bias in general. For the remainder of this report we set the default solution to zero, which is reasonable since most astronomical images are largely composed of pixels corresponding to blank sky \citep{BL06}.

The solution to this system can also be found by considering the singular value decomposition (SVD) of a matrix {\boldmath$B$}$=${\boldmath$U \Sigma$}{\boldmath$V$}$^T$, where {\boldmath$U$} and {\boldmath$V$} are orthogonal $N \times N$ matrices \citep{golub}. The matrix {\boldmath$\Sigma$} is diagonal, containing the non-increasing singular values $\nu_1 \geq \nu_2 \geq ... \geq \nu_n$. The columns of {\boldmath$U$} are a set of orthogonal vectors {\boldmath$u$}$_i$, and the orthogonal columns of {\boldmath$V$} are denoted by {\boldmath$v$}$_i$, which leads us to the expression
\begin{equation}
\mbox{\boldmath$x$}=\sum_{i=1}^{N} \frac{\mbox{\boldmath$u$}_{i}^{T}
\mbox{\boldmath$b$}}{\nu_{i}} \mbox{\boldmath$v$}_i \label{SVD1}.
\end{equation}

\begin{figure}
\scriptsize \epsscale{1} \plotone{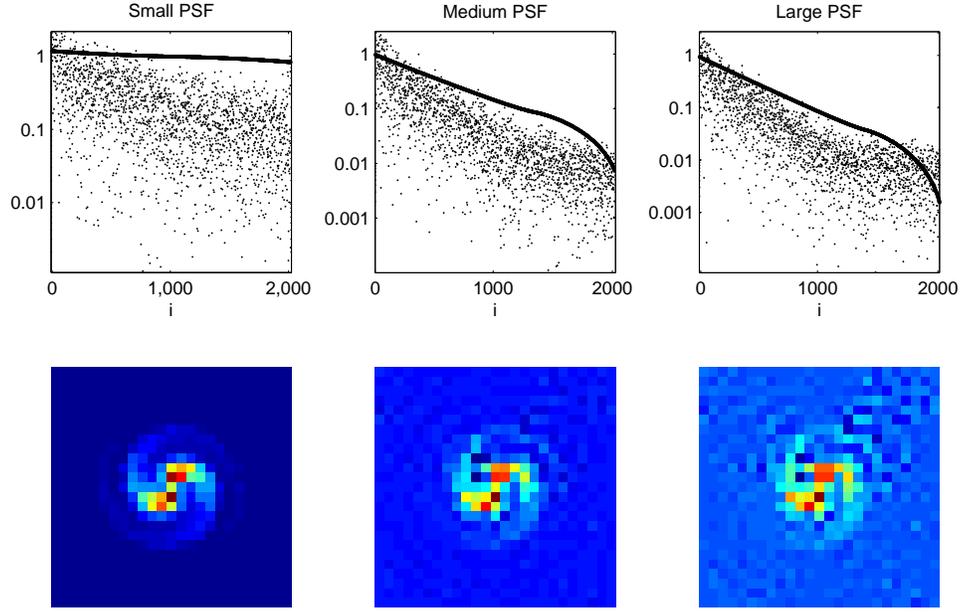} \caption{ Picard plot for Gaussian PSFs with full width at half-maximum of $0.94$, $1.64$, and $2.12$ pixels, respectively. The points on the black curve are the singular values $\nu_i$ and the small dots are the expansion coefficients $|u_i^Tb|$. As the PSF becomes increasingly large, the singular values drop below the expansion coefficients more quickly. This drop-off signifies an increased contribution of high frequency noise in the solution as can be seen in the model solutions shown. } \label{picardPlot}
\end{figure}

The small singular values (large values of $i$) correspond to the addition of high frequency noise, and the terms involving the smallest singular values $\nu_i$ tend to dominate the solution. The singular values $\nu_i$ and the expansion coefficients $|${\boldmath$u$}$_i^T${\boldmath$b$}$|$ as a function of the number of terms are shown in Figure \ref{picardPlot}. These plots are called Picard plots and show that an increased contribution to the noise in the reconstruction is found as the singular values become smaller than the magnitude of the expansion coefficients.

The goal of a regularization scheme is to limit the amount of noise that contributes to the solution. In principle, the simplest scheme is to truncate Equation (\ref{SVD1}) for sufficiently large values of $i$ to limit the amount of high frequency noise in the solution. Truncation of the SVD expansion can be accomplished by multiplying the terms of Equation (\ref{SVD1}) by a ``filter factor'' $\phi_i$ \citep{vogel2} defined by
\begin{equation}
\mbox{\boldmath$x$}_{filt}=\sum_{i=1}^{N} \phi_i \frac{\mbox{\boldmath$u$}_{i}^{T}
\mbox{\boldmath$b$}}{\nu_{i}} \mbox{\boldmath$v$}_i \label{SVD2},
\end{equation}
where $\phi_i$ takes the form of a Heaviside function such that $\phi_i=1$ for singular values below the cutoff point $k$, and $\phi_i=0$ for terms with $i > k$. In this way, the contribution of high-frequency noise to the solution can be controlled. However, this regularization scheme is somewhat artificial because of the sharp cut off in the filter factors. A more natural scheme was developed by \citet{Tik}, which introduces a regularization parameter $\lambda$ to solve
\begin{equation}
\left(\mbox{\boldmath$B$}^T\mbox{\boldmath$B$}+\lambda \mbox{\boldmath$I$}\right)\mbox{\boldmath$x$}
=\mbox{\boldmath$B$}^T\mbox{\boldmath$b$}.
\end{equation}
The Tikhonov solution for $\mbox{\boldmath$x$}$ is expressed as the standard SVD expansion with modified filter factors
\begin{equation}
\phi_i=\frac{\nu_i^2}{\nu_i^2+\lambda} \label{tikFF}.
\end{equation}
The solution of this system corresponds to the solution of Equation (\ref{linsys}) with the regularization matrix equal to the identity and the prior solution \mbox{\boldmath$x$}$_0=$\mbox{\boldmath$0$}.

When $\nu_i \gg \lambda$, $\phi_i \approx 1$. For large $i$, $\nu_i \ll \lambda$ such that $\phi_i \approx \nu_i^2 / \lambda$. Note that the eigenvalues of the $N_s \times N_s$ system matrix $\mbox{\boldmath$B$}^T\mbox{\boldmath$B$}$ are the squares of the singular values, $\mu_i=\nu_i^2$. The sum of the Tikhonov filter factors is then
\begin{equation}
\gamma=\sum_{i=1}^{N_s} \frac{\mu_i}{\mu_i+\lambda}
\label{ndof}.
\end{equation}
This expression agrees with Equation ($21$) in \citet{suyu06}, who show that the sum $\gamma$ represents the number of source degrees of freedom in the problem when Tikhonov regularization is included.

Iterative methods effectively add consecutive terms to the sum in Equation (\ref{SVD1}) with each step of the algorithm, such that the number of iterations itself acts as a regularization parameter \citep{hanke}. In order to find the best solution from an iterative optimizer, it is necessary to stop it near the optimal iteration, before the contributions due to noise in the solution grow too large. As can be seen in Figure \ref{convergeSrc}, the CGLS algorithm converges significantly faster than the SD method. However, the noise also rises more quickly past convergence, which makes the CGLS solution more sensitive to the stopping condition.  The SD method is generally considered a slower and less sophisticated local optimization algorithm than CGLS, but performance issues are outweighed by SD's more stable behavior past convergence.  \citet{nagyPalmer} first noted that optimization schemes based on SD do not require as precise a stopping criterion as other methods, which makes it easier to find an approximation to the optimally regularized solution.  In the absence of the L-curve criteria, we choose SD. When using a stopping condition based on the L-curve, CGLS is recommended due to the algorithms speed in obtaining better solutions.

\subsection{Monte Carlo Estimate of the Effective Degrees of Freedom}
\label{sec:MCDOF}
The iterative optimizers we have considered in this paper can be expressed in terms of the SVD expansion, Equation (\ref{SVD2}), with unique expressions for the filter factors $\phi_i$. As in the case of Tikhonov regularization, we associate the sum of these filter factors $\gamma$ with the number of effective degrees of freedom in the problem \citep{vogel}. For the case of the CGLS algorithm, these filter factors are recursive in the singular values \citep{hansen97}. This poses a problem because we use the CGLS scheme without explicitly building the matrices, and the solution of the singular values presents difficulty when using large data sets. Furthermore, the recursive scheme can become unstable \citep{regu}. To circumvent this problem, we use a Monte Carlo scheme to estimate the sum of the filter factors. In essence, this scheme introduces a Gaussian random vector \mbox{\boldmath$\hat{b}$} with zero mean and unit standard deviation that contains the same number of elements as the data vector \mbox{\boldmath$b$}. While iteratively solving for the solution vector $x$ using the conjugate gradient method, we simultaneously solve a second system with noise vector \mbox{\boldmath$\hat{b}$} using the same CGLS coefficients ($\bar{\alpha_k}$ and $\bar{\beta_k}$ in standard notation) to derive a corresponding vector \mbox{\boldmath$\hat{x}$}. \citet{hanke2} and \citet{girard} show that \mbox{\boldmath$\hat{b}$}$^T($\mbox{\boldmath$\hat{b}$}$-$\mbox{\boldmath$A$}\mbox{\boldmath$\hat{x}$}$)$ provides an estimate of the number of degrees of freedom in the original system with data vector \mbox{\boldmath$b$}.  Note, however, that this estimate approximately doubles the computational overhead of the standard CGLS method.  We perform this calculation during each call to the CGLS algorithm, allowing an estimate of the reduced $\chi^2$ for each set of lens parameters.

\subsection{L-curve Analysis}
\label{sec:Lcurve}

\begin{figure}
\scriptsize \epsscale{1} \plotone{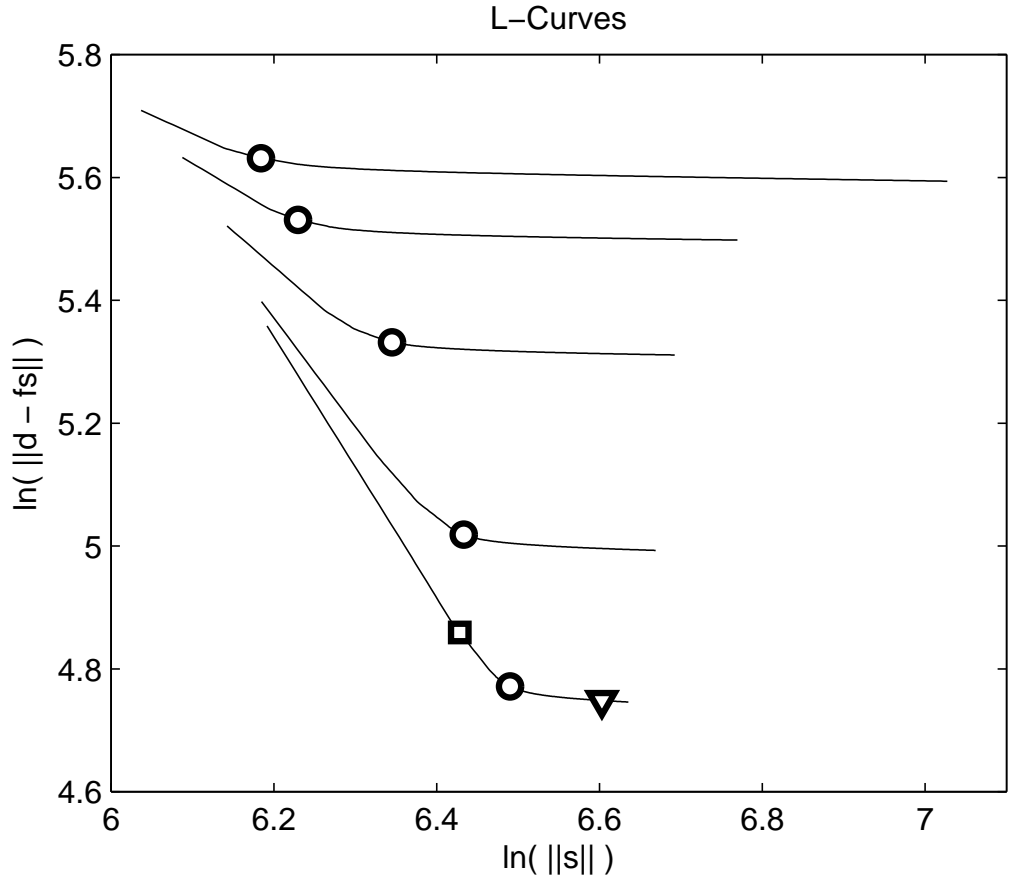} \caption{Variety of L-curves for the system shown in Figure \ref{compareImg}. The curve marked with the square, circle and triangle is the true solution, with the parameters described in Section \ref{sec:SmallTest}. Each successive curve has the same parameters as the true solution except for the velocity dispersion, which takes the values $260$, $262.5$, $265$, $267.5$ and $270$ km s$^{-1}$, respectively. The location of the optimally regularized solution balances the residual and solution norms, denoted by the corner of the curve and marked by a circle. The optimally regularized solution for the true set of lens parameters has reduced $\chi^2=0.998$, $||s||=658.4$, found after $7$ CGLS iterations.} \label{LCurveFig}
\end{figure}

\begin{figure}
\scriptsize \epsscale{1} \plotone{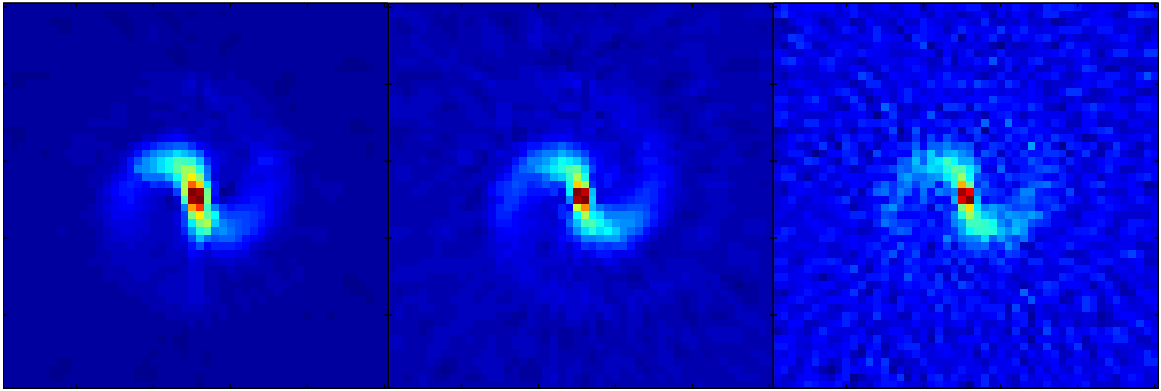} \caption{ Sources corresponding to the solutions marked in Figure \ref{LCurveFig}. Left: oversmoothed solution (square), $3$ CGLS steps, reduced $\chi^2_r=1.228$, $||s_i||=613.2$. The middle panel shows the optimally regularized solution in Figure \ref{LCurveFig} (circle) after $7$ CGLS steps, reduced $\chi^2_r=0.998$, $||s_i||=658.4$. The panel on the right shows the solution (triangle) after $18$ iterations, reduced $\chi^2_r=0.965$, $||s_i||=740.4$.} \label{sources_LCurve}
\end{figure}

The iterative optimization algorithms used in the local optimization step (the inner loop of our optimization scheme) converge to lower spatial frequencies faster than higher frequencies, and therefore the high-frequency noise present in the source reconstructions can be suppressed by controlling the number of iterations of the local optimizer.  In general, we wish to find a balance between the image $\chi^2$ and the amount of source regularization \citep{press}.  Since the regularizing effect of iterative optimizers is implicit, we need a metric to evaluate the amount of regularization introduced at each iteration. For simplicity, we use zeroth-order regularization (\citet{WD03}; \citet{suyu06}) in this paper which sums the squares of source pixel intensities, in order to quantify the regularizing effects of the local optimizers. By calculating the image $\chi^2$ and regularization measure, $\sum_{i}^{N_s}s_i^2$ at each iteration of the local optimizer, we can form an L-curve \citep{hansenOLeary} for each solution.  In the standard image deblurring problem, the point associated with the ``corner'' of the L-curve represents the solution that best balances the image fitness and the amount of regularization introduced in modeling the source.  This solution is found by determining the point on the trade-off curve with maximum curvature.  We parameterize the L-curve by arclength $(x(s),y(s))$, where $x$ and $y$ represent the regularization term and image $\chi^2$, respectively, and fit a cubic spline curve to $x$ and $y$.  The derivatives of the cubic spline curves with respect to the arclength can be calculated analytically, which provide a simple method to calculate the curvature $\kappa$. The point of maximum curvature is found using the curvature formula:
\begin{equation}
\kappa=\frac{|x'y''-y'x''|}{\left( x'^2+y'^2 \right)^{\frac{3}{2}}}.
\end{equation}
We show a sample of L-curves in Figure \ref{sources_LCurve} and corresponding source solutions in Figure \ref{LCurveFig}, including the solution located at the point of maximum curvature. In general, the solution found by the L-curve analysis agrees with the maximum Bayesian evidence solution to approximately 10\%. The solution corresponding to the point of maximum curvature of the L-curve is used to evaluate the fitness of each set of lens parameters.

\subsection{A Large-scale Test}
\label{sec:LargeScaleTest}
\begin{figure}
\scriptsize \epsscale{0.75} \plotone{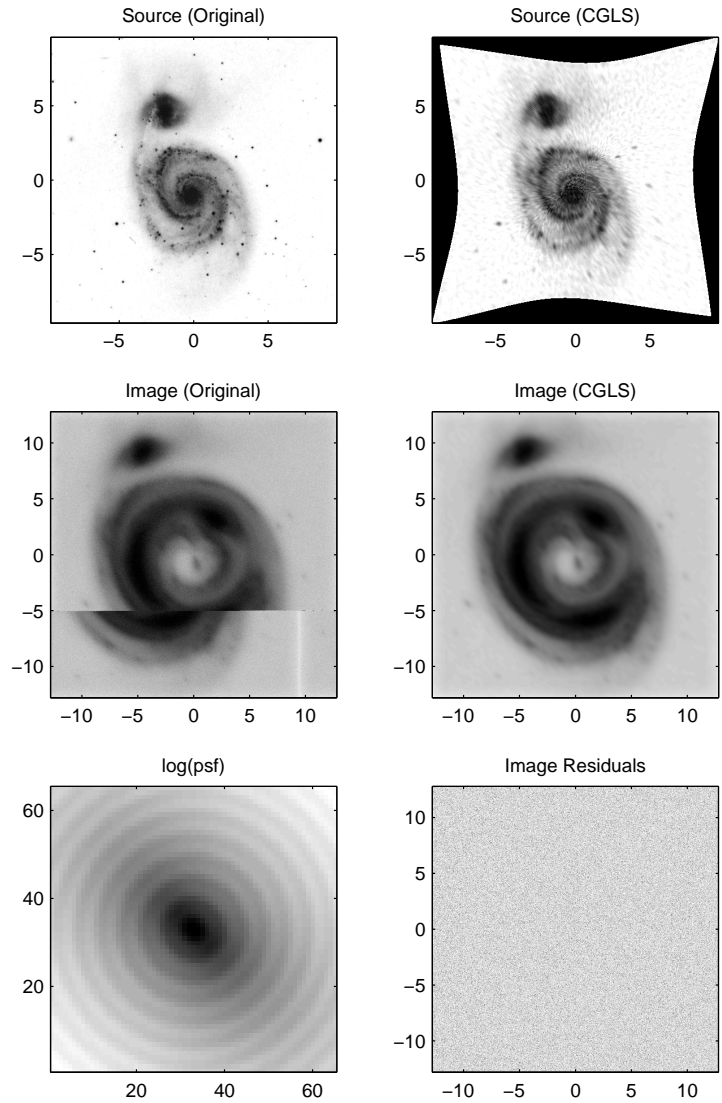} \caption{Large scale test of Mirage. Top left: original image of $M51$ used to generate a large-scale test problem. Top right: model source obtained with the CGLS algorithm after $40$ iterations. The original image was obtained from NED, originally $700 \times 700$, cropped to $512 \times 512$ pixels. The effect of the lens mapping can be seen as the source plane is not completely covered by the back-traced image. Middle left: lensed image of $M51$ as seen through an SIE lens used as artificial data. The image is comprised of $640 \times 640$ pixels, over an area of 25.5 arcsec$^2$. Middle right: model image of $M51$ as produced by the CGLS algorithm after $40$ iterations. Bottom left: PSF used to blur the observation, shown in logarithmic intensity to highlight the low-level structure. The function is a $65 \times 65$ pixel PSF. Bottom right: residuals obtained from comparing original and model images. The residuals are featureless and have a maximum $10^{-3}$ of the original image maximum. The reconstruction has a reduced image $\chi^2_r = 0.9954$.} \label{M51Fig}
\end{figure}

We form the gravitationally lensed image of a large source to demonstrate the efficiency of our iterative matrix-free approach. The source is a square image of $M51$, of dimension $512 \times 512$, shown in Figure \ref{M51Fig}, obtained from the NED online data archive \citep{kennicut}. The lensed image is generated using an SIE lens defined on a $640 \times 640$ grid, with Einstein radius $b$=3 arcsec, $\epsilon=0.4$, and $\theta_L=\pi/4$, with the lens centered at the origin. To demonstrate the behavior of the code with a complicated PSF, we used a PSF composed of a radial sinc function multiplied by an elliptical Moffat PSF \citep{moffat}, which is shown in the figure. The resulting function provides a non-symmetric PSF that contains significant low-level structure. Such a large PSF would require a very large non-sparse blurring matrix, whose linear size must necessarily match the number of image pixels which map to the source, in this case $3.34 \times 10^5$ square. After adding Gaussian white noise, the S/N of the blurred observation is $S/N=20$. The solution shown in the figure was computed by the CGLS method and has a reduced image $\chi^2_r = 0.9954$ and was found in $35$ iterations that took $86.4$ seconds using a single $2.4$ GHz Intel processor.

The next section discusses the global optimizers, Ferret and Locust, which we use to solve for the lens model parameters.  Both are parallel codes that require approximately $10^4-10^5$ lens parameter sets to be evaluated for a thorough search, optimization, and mapping of the parameter space.  Assuming $5\times 10^4$ evaluations, the lens parameters could be solved for this large-scale test problem in approximately six days on an eight-core computer.  Such a large-scale problem would be impractical using a matrix inversion scheme due to the large size of the matrices that would be involved.

\section{The Full Optimization Problem}
\label{sec:fullopt}
Section \ref{sec:lensing} focused mainly on the linear least squares reconstruction of the source, for a known lens mass distribution.  However, the full problem must also determine the optimal set of lens parameters.  The lens parameters are solved as an `outer loop' optimization problem, which calls the semilinear method, or alternatively our iterative approach, as an inner loop optimization for each set of nonlinear lens parameters evaluated.  The inner loop optimizes the lensed source by executing an arbitrary number of iterative steps (in our examples, $40$) of a local optimizer like the CGLS algorithm. The L-curve for each lens parameter set is built, and the optimally regularized solution that lies nearest the corner of this curve is found. The $\chi^2$ value of this optimally regularized solution is used to evaluate the fitness of the corresponding set of lens parameters.  During the inner loop optimization, a statistical estimate of the number of degrees of freedom for the optimally regularized solution is made and used to determine the reduced image $\chi^2$ during the analysis at the end of the run.  In this paper, the outer loop problem is solved by the Ferret GA and Locust PSO from the Qubist Global Optimization Toolbox \citep{Fiege}.   However, the Mirage code is not limited to either of these optimizers and can make use of any external nonlinear optimization scheme.

Both Ferret and Locust are able to map out ``fuzzy'' optimal sets defined by an inequality.  In this case, we request a distribution of solutions with $\chi^2 \le \chi^2_{min}+N_u$, where $\chi^2_{min}$ is the lowest image $\chi^2$ value found and $N_u$ is an upper limit selected at the start of the run.  The upper limit $N_u$ is chosen to be large enough so as to include solutions within the $99\%$ confidence interval.  The members of the optimal set, along with the estimates for the number of degrees of freedom, allow us to determine solutions within the $99 \%$, $95 \%$ and $68 \%$ confidence intervals by the standard method \citep{press}.  Thus, we can easily estimate errors for the nonlinear lens parameters, since these global optimizers determine the form of the $\chi^2$ surface in the neighborhood of the global minimum.

The source intensities may contain negative values since bounds cannot be imposed in direct matrix inversion, and are not enforced in our iterative schemes.  A final source refinement step, discussed in Section \ref{sec:sourceRefine}, uses the GA and PSO as bounded optimizers to find the optimal positive definite source distribution, with the lens parameters held fixed at their previously optimized values.

\subsection{Global Nonlinear Optimization}
\label{sec:global}
The Qubist Global Optimization Toolbox contains five global optimizers in total, all of which are designed to be interchangeable.  Ferret and Locust are the most powerful and well-tested optimizers in the package, which makes them well-suited for our problem.  Qubist includes more than $50$ test problems, some of which are discussed in its user's guide \citep{Fiege}.

GAs and PSOs differ greatly from local optimization routines such as CGLS and SD, which require an initial guess and then search iteratively along a deterministic trajectory through the parameter space.  Such methods are prone to becoming trapped in local minima.  Moreover, these methods are usually implemented to solve unbounded optimization problems, which may be less useful than bounded optimization when there are physical constraints on the parameters, such as the positivity of source pixel values in the lens reconstruction problem (see Section \ref{sec:SL}).

GAs and PSOs search the parameter space in parallel, making use of the collective behavior of numerous interacting ``agents'' - a population of individuals for a GA or a swarm of particles in the case of a PSO.  These optimizers distribute agents randomly throughout the parameter space initially, which subsequently interact using heuristic rules that aim to search the space thoroughly, and encourage the improvement of the population or swarm as a whole.  In both types of algorithm, these heuristic rules are partly deterministic and partly stochastic.  The resulting optimization algorithms are more powerful and robust than purely deterministic methods and vastly more efficient than random search.  In general, only a single agent must find the high-performance region in the vicinity of the true global solution for the algorithm to succeed.  Once such a solution is discovered, it is rapidly communicated to all other individuals or particles, which will accumulate near the global minimum and refine it.

\subsubsection{Genetic Algorithms}
\label{sec:GA}
GAs are an important class of algorithms for global optimization that work in analogy to biological evolution.  Evolution is biology's optimization strategy of choice, in which organisms evolve and continually improve their own designs as they struggle to survive.  GAs are normally discussed using biological terminology, such that each ``individual'' is a trial solution, whose parameters are encoded on ``genes''.   The set of individuals is a ``population'', and individuals search the parameter space in parallel as they evolve over multiple ``generations''.  A basic GA requires three genetic operators, which are mutation, crossover, and selection \citep{goldberg89}.  The role of mutation is to apply occasional random perturbations to individuals, which helps them to explore new regions of the parameter space.  Crossover mixes together two parent solutions to produce offspring that are intermediate between the parent solutions.  The role of the selection operator is to choose which solutions propagate to the next generation, based on the Darwinian notion of survival of the fittest.  Various types of selection operators are possible, but tournament selection has the advantage that it is insensitive to the scaling of the fitness function \citep{goldberg02}.

Ferret is a parallel, multi-objective GA, which has been under constant development since 2002, and is the most sophisticated optimizer in the Qubist package. The current version is the fourth major version of the code, and earlier versions were used by \citet{fiege04} to model magnetized filamentary molecular clouds, and by \citet{fiege05} to model submillimeter polarization patterns of magnetized molecular cloud cores.  Ferret extends the basic GA paradigm in several important ways, as discussed below.

Multi-objective optimizers like Ferret emphasize the thorough exploration of parameter spaces and the ability to map trade-off surfaces between multiple objective functions, which allows the user to understand the compromises that must be made between several conflicting objectives.  A core feature of a multi-objective GA is the ability to spread solutions approximately evenly over an extended optimal set of solutions, which Ferret accomplishes using a niching algorithm similar to the one discussed by \citet{fonseca93}.  Even for single-objective problems, Ferret's multi-objective machinery is well-suited to explore and map out $\chi^2$ intervals in the neighborhood of the global minimum.  We see in Section \ref{sec:results} that it is especially useful for degenerate cases where multiple disconnected islands of solutions exist within the parameter space.

Ferret's most novel and powerful feature is its `linkage-learning' algorithm \citep{goldberg02}, which is designed to reduce a complex, multi-parameter problem to a natural set of smaller sub-problems, whenever such a reduction is possible.  These simpler sub-problems are discovered experimentally by Ferret during the process of optimization, and sub-problems evolve almost independently during a run.  Ferret regards two parameters $A$ and $B$ as linked if finite variations of $A$ and $B$ are discovered, which result in worsening of a solution when applied independently, but the same variations applied together result in improvement.  In such cases, it is clear that $A$ and $B$ should be linked so that they are usually traded together during crossovers, to preserve gains made by varying the parameters together.  A novel extension of Ferret's linkage-learning algorithm is its ability to search entire sets of parameters $\{A_i\}$ and $\{B_i\}$ for linkage in parallel, which is assigned probabilistically to the parameters within these sets.  Thus, Ferret treats linkage as a matrix of probabilities that co-evolves with the population during the search.  Parameters that appear linked at the start of a run may not appear linked at the end, when most solutions may be nearly optimal.  Conversely, new links can also arise as the code explores previously uncharted regions of parameter space.

The ability to partition a complicated problem into natural sub-problems is crucial to the successful optimization of large problems.  A difficult 100 parameter problem with many local minima is often unsolvable on its own, but it becomes quite tractable if it can be partitioned into (say) $10$ sub-problems (or building blocks) with $10$ parameters each.   A particularly interesting feature of Ferret's linkage-learning system is that the linkages discovered are entirely insensitive to scale.  Two sub-problems (building blocks) that are orders of magnitude different in importance are discovered at the same rate, so that Ferret can solve all of the sub-problems correctly and simultaneously, rather than one at a time in order of significance.  This ability allows Ferret to discover the true, globally optimal solution or solution set, even when applied to problems with very poorly scaled building blocks.

Ferret contains an algorithm that monitors its progress and uses this information to automatically adapt several of its most important control parameters, including the mutation scale, size scale of crossover events, and several others.  If these parameters are set poorly by the user, Ferret quickly and dynamically adapts them to improve the search.  This algorithm provides an extra layer of robustness to the code, which helps Ferret to adapt as different regions of the fitness landscape are discovered.

Ferret, and the other global optimizers of the Qubist toolbox, place considerable emphasis on visualization.  The analysis window displayed at the end of a run contains various graphics options to tease out interesting features from the optimal set.  These features include two and three-dimensional scatter plots, image plots, contour plots, and user-defined graphics.  It is possible to `paint' interesting regions of the parameter space and select different two and three-dimensional projections to explore and visualize where the painted solutions reside in a high-dimensional parameter space.

Modeling a gravitational lens system is a computationally intensive task that requires approximately $10^4-10^5$ parameter sets to be evaluated for a single run.  GAs are well suited to parallel computing because each individual in the population represents a single parameter set, which can be evaluated independently.  Ferret is designed with built-in parallelization to take advantage of multi-CPU computers and inexpensive clusters.  Parallel jobs are managed with a graphical ``node manager'' tool, and no changes are required to the implementation of the user's fitness function.  It is notable that Ferret does not require MATLAB's parallel computing toolbox, or use any other third-party parallel computing software.

The Appendix discusses some additional details of the Ferret algorithm.

\subsubsection{Particle Swarm Optimizers}

Locust is a parallel multi-objective PSO in the Qubist toolbox.  PSOs are biologically inspired global optimizers, which search the parameter space using a swarm of interacting particles.  PSOs are often discussed in terms of the dynamics of flocks of birds, schools of fish, or swarms of social insects searching for food.  The commonality is that intelligent search behavior emerges as property of the system as a whole, even if the component parts are modeled as relatively simple automata that interact with each other through simple rules.  \citet{eberhart01} provides a good introduction to the PSO technique.

PSOs are similar to GAs in that they sample many points in the search space simultaneously, with a swarm of particles moving through the parameter space following simple dynamical equations.  Each particle in a simple PSO is simultaneously attracted to its own ``personal best'' solution, which is the best solution that the particle has personally discovered, and the ``global best'' solution, which is the best solution that the entire swarm has ever encountered.  The law of attraction follows a simple spring law: $F \propto |\Delta {\bf x}|$, where $|\Delta {\bf x}|$ is the distance between a given particle and either the personal best solution ${\bf x}_p$ or the global best ${\bf x}_g$.  Assuming that the force and velocity are approximately constant over a time step, the new velocity and position of particle $i$ after a time step $\Delta t$ are given by
\begin{eqnarray}
{\bf v}_i(t+\Delta t) &=& {\bf v}_i(t)\left(1-\Delta t/t_{damp}\right) + \nonumber\\
&&{} \left[c_p\xi_p({\bf x}_p-{\bf x}_i)+c_g\xi_g({\bf x}_g-{\bf x}_i)\right] \Delta t \nonumber\\
{\bf x}_i(t+\Delta t) &=& {\bf x}_i(t)+{\bf v}_i(t) \Delta t,
\label{eq:PSO}
\end{eqnarray}
where $c_p$ and $c_g$ play the role of spring constants for the personal and global best solutions respectively.  The equations include a damping term to decrease the velocity magnitude in approximately time $t_{damp}$, which helps the swarm settle down as it zeros in on the optimal region.  Damping also serves to prevent runaway growth in so-called `particle explosions', which can occur as a result of accumulated errors in Equation (\ref{eq:PSO}).  Some randomness is added via the uniform random variables $\xi_p$ and $\xi_g$, which are typically drawn from the range 0-1.  The stochastic terms play a role similar to the mutation operator in a GA; they add randomness to the search, which helps the particles to explore previously unexplored parts of the parameter space.  The roles of the personal and global best solutions are clear. The personal best solution represents a particle's memory of the best region of parameter space that it has seen, and the global best solution represents the entire swarm's collective memory.  In effect, the global best solution allows indirect communication between particles to encourage collective behavior.

Particle swarm optimization is a young and rapidly changing field of research that still has many open questions, which are discussed in a recent review by \citet{poli}.  Equation (\ref{eq:PSO}) is perhaps the simplest set of swarm equations, but many alternative implementations are possible, which strive to balance thorough exploration of the parameter space against the need to exploit high performance regions when they are found.

Equation (\ref{eq:PSO}) is equivalent to a simple Euler integration scheme for a dynamical system of equations that move each particle every time step.  However, Locust uses an exact solution to the swarm equations, which is easily obtained by solving Equation (\ref{eq:PSO}) analytically, in the limit $\Delta t \rightarrow 0$.  Numerical experiments with Locust, and alternate schemes that use Euler integration, show that the exact solution results in more stable and reliable PSO \citep{Fiege}.  It is possible that the exact solution eliminates the build-up of errors in the orbits, which would result from applying Equation (\ref{eq:PSO}) directly with a finite $\Delta t$.  The exact solution is slightly more costly to evaluate than the Euler approximation, but this extra computational expense is insignificant for any realistic problem, where the computational time is normally dominated by the evaluation of the fitness function.

Determining ${\bf x}_p$ is straightforward because it represents the personal best solution (often denoted {\em pbest}) that any particle has encountered.  Thus, each particle simply keeps track of the position where it encounters the lowest value of the fitness function $F({\bf x}$), following Ferret's convention that lower values of $F$ correspond to more desirable solutions.

The most common particle swarm implementation is the simple PSO discussed above, where the global best solution ${\bf x}_g$ ({\em gbest}) is evaluated over the entire swarm.  This swarm topology can be thought of as a fully connected graph, where each particle in the swarm communicates with every other particle via the {\em gbest} solution.  Other swarm topologies are possible, where the network of communication between swarm members is less densely connected, so that each particle only communicates with a few other particles in its neighborhood.  In this case, the {\em gbest} solution is replaced by a set of local best, or {\em lbest} solutions, such that each {\em lbest} solution is assigned to a subset of the swarm.  This scenario is referred to as a static {\em lbest} topology when the network connecting particles do not change throughout the run.  Dynamic {\em lbest} topologies are also possible, where the network co-evolves with the swarm as the run progresses.  Swarms based on sparsely connected networks can be thought of as being divided into sub-swarms, where each sub-swarm shares a common {\em lbest} solution.  Such a topology is better able to avoid local minima because the sub-swarms explore the space in parallel.  On the other hand, the fully connected {\em gbest} topology is best for exploiting a single isolated solution late in a run, because it focuses the efforts of the entire swarm on the region of parameter space in the vicinity of the {\em gbest} solution.

Locust requires some non-standard techniques designed to thoroughly explore parameter spaces containing sets of solutions that are equally good.  Extended solution sets are also possible when a fuzzy tolerance is specified for a single objective problem, which often represents the $\chi^2$ error tolerance of a data-modeling problems.  Locust emphasizes the mapping of spatially extended solution sets, and therefore it makes sense to define particle neighborhoods dynamically, based on their spatial location within the swarm.  The code keeps track of the Euclidean distances between all particles, and assigns neighborhoods based on the nearest {\em lbest} particle.  Moreover, Locust implements a novel algorithm that allows neighborhoods, and hence sub-swarms, to merge and divide as required to map out the structure of the optimal set.  This dynamic swarm topology is quite different from other topologies discussed in the literature, and has the benefit that it essentially self-optimizes.  A large number of neighborhoods will generally be preserved to map a spatially extended solution set, but the swarm topology will correctly collapse to a single neighborhood late in a run if only a single solution exists, thus reducing the algorithm to a simple {\em gbest} approach.  In practice, this technique represents a good balance between exploration of the parameter space and exploitation of the optimal set; the parallel action of many sub-swarms evade local minima early in the run for all problems, and many are retained to the end when the focus is on mapping an extended solution set, but swarms reduce to the maximally exploiting {\em gbest} algorithm late in the run for problems where only a single best solution exists.

Locust uses the same visualization system as Ferret.  It uses a simpler setup file than Ferret, but it can read Ferret's setup files and translate them.  Moreover, the formats for the initialization, fitness, and custom graphics functions are identical.  This makes it easy to swap optimizers for comparison purposes. The Appendix discusses some additional details about Locust.

\subsection{Source Refinement Routine}
\label{sec:sourceRefine}
We use a two-step process to solve the full inversion problem. In the first step, we determine the nonlinear lens parameters as described in Section \ref{sec:fullopt}. In the second refinement step, we hold the best set of lens parameters constant and allow the global optimizer to fit the source brightness distribution. We treat each source plane pixel as a free parameter and judge the fitness of solutions based on the image $\chi^2_r$ statistic.  This type of pixelized source fitting using a GA was outlined by \citet{BLGA}. The Qubist global optimization routines are bounded, so positivity conditions on the source reconstruction are easily enforced in this step. Since the intensity of each source pixel is independent, this approach does not produce a regularizing effect and the number of degrees of freedom in the problem is well defined, allowing direct comparisons of lens models.  Therefore, this two-step method allows an estimation of the errors on both the lens and source intensity parameters.

\begin{figure}
\hspace*{-0.3\textwidth}
\scriptsize \epsscale{2} \plotone{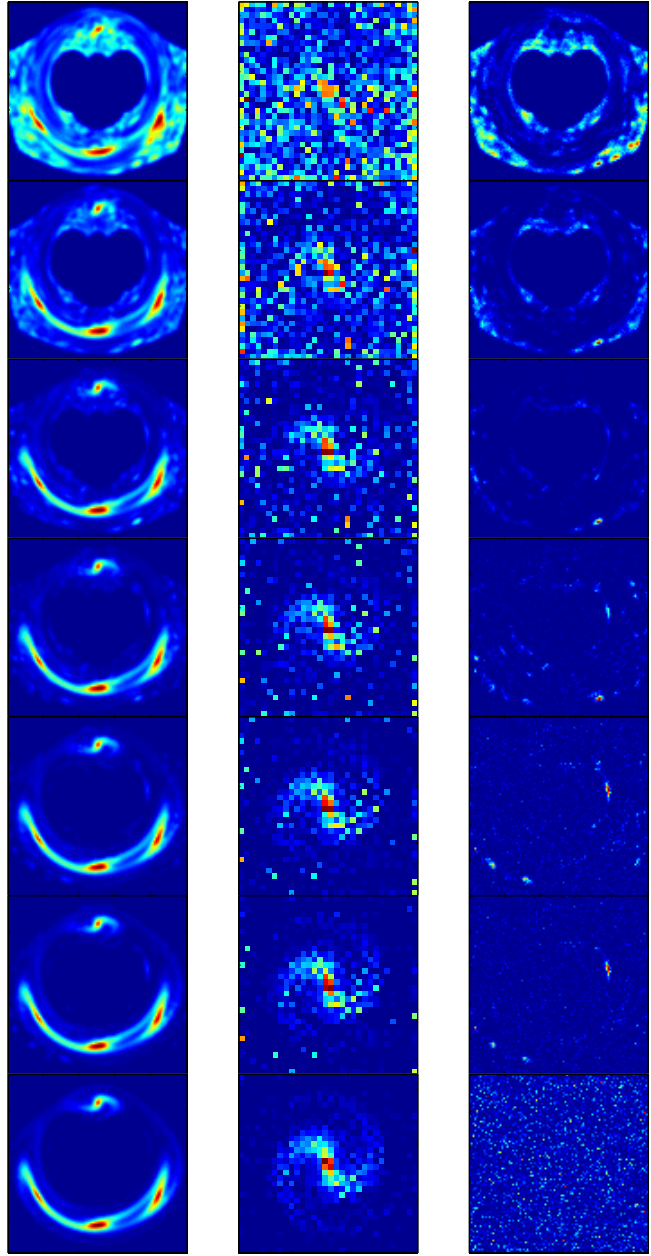} \caption{ The lowest $\chi^2_r$ solution at $100$, $250$, $500$, $750$, $1000$, $1250$, $5000$ generations. Left column: model image. Middle column: source brightness distribution. Note the presence of reconstructed noise.  Right column: image residuals. At $5000$ generations, a model image with $\chi^2_r=1.05$ was found. Each image is independently scaled to highlight image features.} \label{BLplot}
\end{figure}

The bounds used in the refinement step can significantly speed up this optimization. Figure \ref{BLplot} shows a sequence of solutions with a lower bound of $0$ and an upper bound equal to $1.1$ times the maximum pixel intensity in the source. These bounds ensure that the source is strictly positive but can significantly slow the optimization due to the large volume of the parameter space that is searched. Both of the global optimizers used in this report can include a user-defined solution in the first generation. Therefore, a more practical optimization strategy is to consider the absolute value of the optimally regularized solution found by the iterative optimization process, and define a ``window'' of acceptable pixel intensity values for each source pixel. In our tests, a tolerance of $\pm 25 \%$ of the pixel intensities is usually sufficient to bracket the true intensities. Pixels with negative intensities in the optimally regularized solution should always have a lower bound of $0$ to prevent artifacts in the source solution. The upper bound of these pixels is taken to be the absolute value of the pixel intensity plus $15\%$. Practically, this reduces the volume parameter space to be searched and generally allows a solution to be found more quickly.

\section{Demonstrations of the Full Optimization Problem}
\label{sec:results}
In this section, we show results from several illuminating test problems that solve the full lens reconstruction problem and characterize the behavior, performance, and limitations of the global optimizers.

\subsection{Trivial Solutions and the Problem of Dimensionality}
Consider a lens model based on a singular isothermal sphere, which provides a simple analytical model with a circularly symmetric deflection angle given by Equation (\ref{einsteinRadius}) in the radial direction.  This deflection angle is used to form the synthetic data with velocity dispersion $\sigma_v=500$ km s$^{-1}$, centered at the origin $\left( x,y \right)=(0,0)$.  We construct artificial data where the Einstein ring has radius $b=1$ arcsec, assuming source redshift $z_d=0.2$, deflector redshift $z_s=1.5$.  For convenience, we again measure angular distances with respect to the Friedmann metric with $k=0$. The lensed image is defined on a $120 \times 120$ rectangular mesh with an image pixel size of $0.015$ arcsec. A $3 \times 3$ subsampling per pixel is used to construct the lensed image. The source is perfectly aligned with the lens center and forms a full Einstein ring due to the symmetry of the mass distribution.  We have blurred the image using a Gaussian PSF with an FWHM of 2.35 image pixels defined on a $33 \times 33$ grid. The test source is also a Gaussian model on a $50 \times 50$ square mesh from $-3$ to $3$ arcsec in both directions.

\begin{figure}
\scriptsize \epsscale{0.9} \plotone{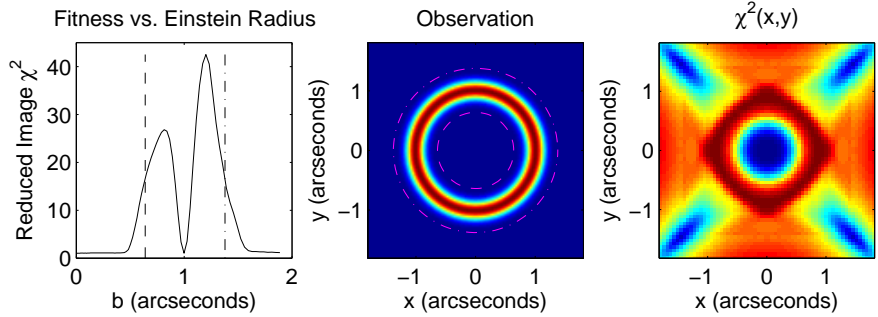} \caption{Left: fitness as a function of Einstein radius for a symmetric lensed image produced by a background Gaussian source intensity distribution.  The Einstein radius was varied manually with the lens center fixed at the origin. The dashed line indicates the lower limit used to model the system, and the dash-dotted line is the upper limit used to restrict the value of the Einstein radius. The region between these two limits is the region in which the true solution is located. Middle: Einstein radius limits are shown superimposed on the artificial data. Right: fitness as a function of lens center. The lens center position was varied over a $64 \times 64$ grid and the lens normalization fixed at the true value, $b=1$ arcsec. Trivial solutions populate the corners of the image. The true solution lies at the center of the image.} \label{SISCutDouble}
\end{figure}

In the following discussion, we hold the $x$ and $y$ coordinates of the lens center constant, using the actual values from the artificial data, and plot $\chi^2_r$ as a function of $b$ in Figure \ref{SISCutDouble}. As the size of the Einstein radius (velocity dispersion) is varied, the corresponding $\chi^2_r$ statistic becomes double peaked, with the true solution between the peaks.  The area to the left of the peaks, the region of low $b$, contains trivial solutions that map the source almost straight through the lens, reproducing the image almost exactly with minimal distortion.  In fact, the $b=0$ source does not include any gravitational lens effect at all, thus reducing the problem to a conventional image deconvolution exercise. Note that this trivial solution results in reduced $\chi^2_r=0.973$, even though the reconstructed source is physically unrealistic.  The $\chi^2$ surface varies smoothly as we approach the `true' value of $b$ with $\chi^2_r=0.986$, and increases with $b$ beyond this value. When $b$ is large, we again begin to see a decrease in $\chi^2_r$, to the asymptotic value of $\chi^2_r=1.15$, as the structure of the source becomes increasingly complex to compensate for the distortion introduced by the lens.  Typically such high $b$ solutions give rise to spurious images in which some pixels lie outside the boundaries of the image plane. We wish to avoid the very low and very high $b$ solutions, since they do not correspond to physical solutions of the problem.

With the lens center fixed, the above example is a simple one parameter problem, which can be easily solved by a global optimization routine designed to map a range of $\chi^2$ values near the global minimum.  However, analogous examples may exist in more complicated systems with more parameters, where the parameter space can become dominated by trivial solutions. The problem becomes especially difficult when false solutions, such as the trivial ones in Figure \ref{SISCutDouble}, occupy a region of space whose dimensionality is greater than the true solutions. In such cases, GAs and PSOs can fail when the number of search agents is too small for the problem, since the entire population or swarm may be drawn into the region of trivial solutions and never discover the class of true solutions that occupy a region of lower dimensionality.  Even if the high-performance region containing the true solution is discovered, both Ferret and Locust are designed to spread solutions evenly over the optimal region, which contains the trivial solutions if the goal is to map the solution set within $\Delta\chi^2$ of the $\chi^2$ minimum.  Thus, the population or swarm may become diluted by spreading out over the trivial region, which has higher dimensionality.  The right panel of Figure \ref{SISCutDouble} shows the results of keeping the Einstein radius $b$ fixed at its true value and varying the lens center $(x,y)$. The true solution point, with $\chi^2_r=0.986$, is surrounded by a ring of poor solutions, which signifies multiply imaged solutions. In this projection, trivial solutions occupy a two-dimensional plane at large radius and have $\chi^2_r=0.978$. In order to overcome the complication of trivial solutions, an estimate of the range of acceptable parameter values is made. By imposing such parameter restrictions the algorithm is guaranteed to find a non-trivial solution to the optimization problem.  Notably, Ferret also implements a novel algorithm that promotes the speciation of the population into isolated clusters, which may help to overcome this difficulty.

\subsection{A Realistic Test}
\begin{figure}
\scriptsize \epsscale{1} \plotone{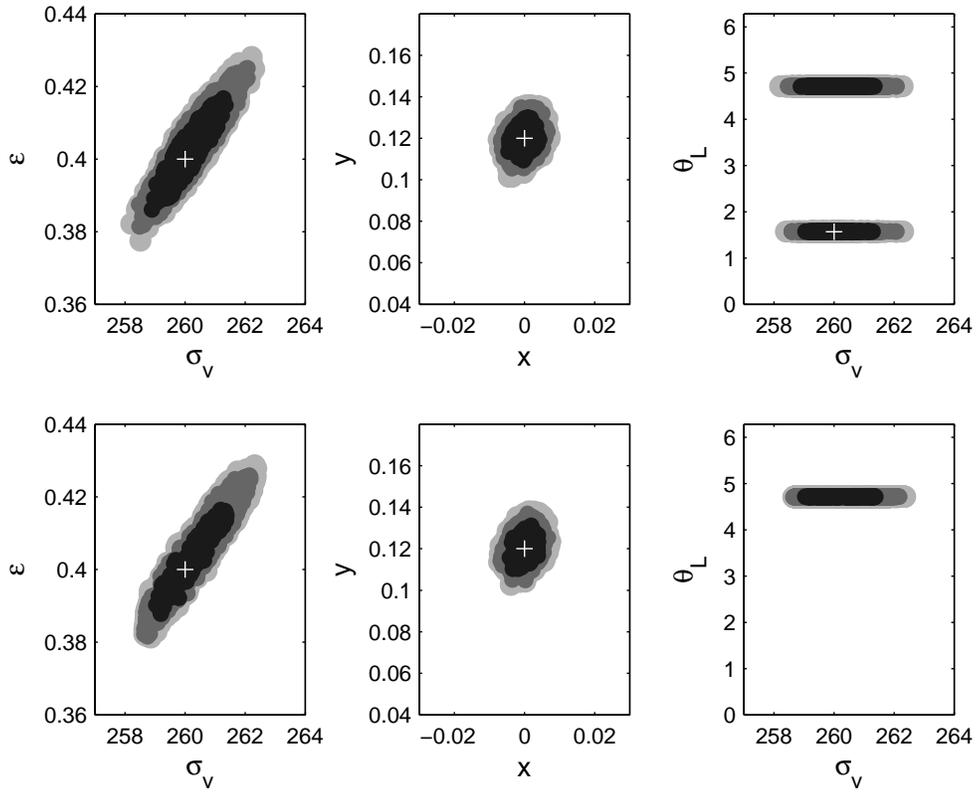} \caption{Parameter space plots of the SIE example in Section $4$. The optimal set of solutions determined by global optimization marked by points shaded according to position within the $99\%$, $95\%$ and $68\%$ confidence intervals, represented by light gray, medium gray and black respectively.  The location of the true solution is marked with a white cross.  Top row: the Ferret GA optimal set. Bottom row: the Locust PSO optimal set. Left column: ellipticity $\epsilon$ vs. velocity dispersion $\sigma_v$, middle column: lens centre coordinates $y$ vs. $x$, and right column: orientation angle $\theta_L$ vs. $\sigma_v$.  The PSO does not explore the structure of the parameter space as thoroughly as the GA. Note the rightmost column in which the orientation angle degeneracy of the system is detected by the Ferret GA but no corresponding solution group is present in the Locust PSO optimal set of solutions.} \label{parameterSpaceSmall}
\end{figure}

For a more realistic and complicated test, consider the SIE lens model presented in Section \ref{sec:SmallTest}. We use the same parameters to solve a test system, with the source intensity profile as given by Equation (\ref{bonner}). We fix the redshift of the deflector and source as in the previous example, and model the parameters of the lens density model using both Ferret and Locust. The fitness objective to be minimized is the standard $\chi^2$. The parameters of the best solutions are summarized in Table $1$.
Both algorithms automatically map the region of parameter space near the
minimum by heavily populating this region of parameter space. The effective number of degrees of freedom for each model is estimated and saved during the course of the run. By using these quantities we are able to estimate confidence intervals and the errors of the lens parameters.  The structure of the global $\chi^2$ surface is calculated at the end of the run using the members of the optimal set, saved from each generation (Ferret) or time step (Locust).  Figure \ref{parameterSpaceSmall} shows that Ferret more thoroughly explores the parameter space than the Locust algorithm.

\begin{figure}
\scriptsize \epsscale{1} \plotone{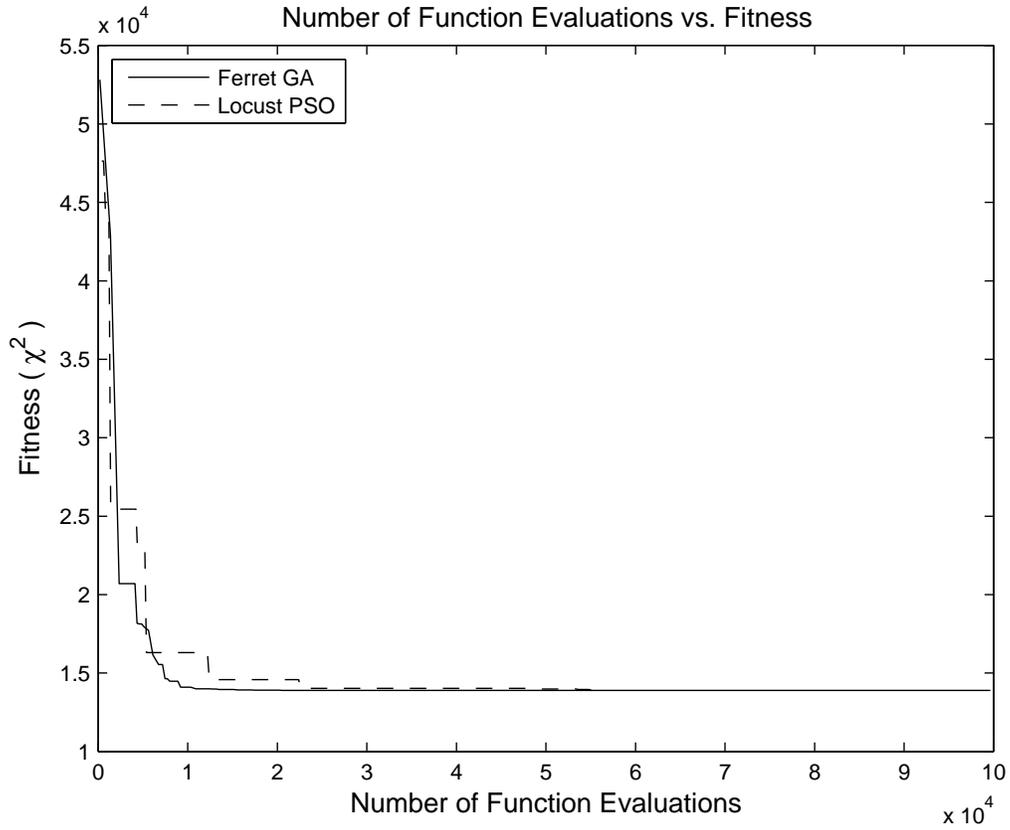} \caption{Average convergence history of the Ferret GA (solid line) and the Locust PSO (dashed line) as a function of the number of function evaluations. The convergence of the GA is more stable, as the PSO tends to converge in a series of steps as the parameter space is explored. The figure is plotted over $1.0 \times 10^5$ function evaluations. We have averaged over four runs of the PSO and four runs of the GA.} \label{convergeGAPSO}
\end{figure}

We find that the GA and PSO converge approximately at the same rate. Figure \ref{convergeGAPSO} compares the performance of the algorithms by plotting the fitness of the best solution as a function of the number of function evaluations, while Figure \ref{parameterSpaceSmall} shows the distribution of solutions in the parameter space. Figure \ref{parameterSpaceSmall} shows that Ferret correctly discovers a pair of equally good degenerate solutions symmetric in orientation angle, but Locust picks out only one of these groups, which reflects Ferret's greater emphasis on mapping the parameter space.  \citet{stasiThesis} used these same optimizers to estimate the system temperature of the DRAO synthesis array and noted that Locust found solutions significantly faster on average.  We do not find the same behavior of the PSO in this problem.

\subsection{Source Refinement}
Once we have determined the lens parameters, we hold them constant and begin the final source refinement step of the optimization, which involves $2500$ parameters for the case shown.  The source refinement results in an optimal non-negative source intensity distribution, as discussed in Section \ref{sec:sourceRefine}. The image is of size $120\times120$, while the source plane is defined as a $50\times50$ grid. The solutions at the beginning of this step appear to be comprised purely of noise, but an approximation to the true solution becomes increasingly well defined as the run progresses, and the image residuals gradually become featureless.  Figure \ref{BLplot} shows an evolutionary sequence of the lowest $\chi^2_r$ solution, where the final solution has $\chi^2_r=1.05$.  The search is a bounded linear problem, which is mathematically simpler than the nonlinear search for lens parameters.  However, the large number of parameters complicates the optimization and the GA converges in a few thousand generations. Source refinement is the most computationally expensive part of this problem, requiring approximately five days on an eight core machine. The most useful aspect of this intensive search is to estimate errors on the source plane pixels determined by the best fit lens model.

\begin{figure}
\scriptsize \epsscale{1} \plotone{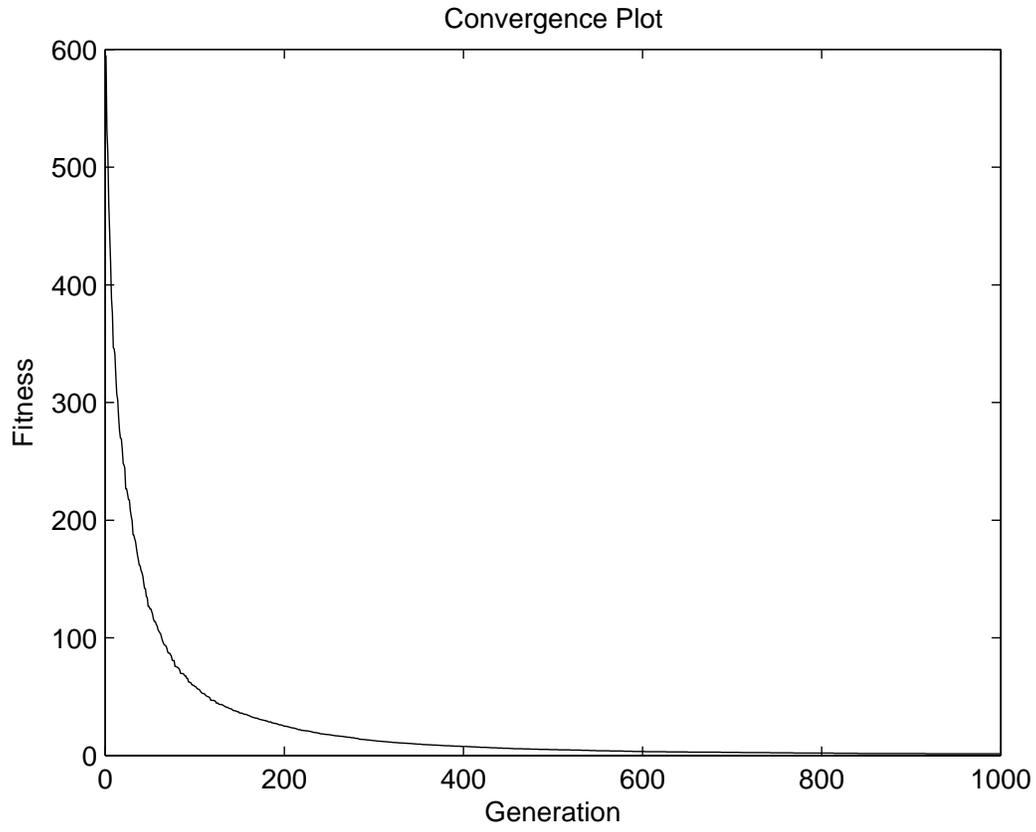} \caption{ Convergence history of the linear parameters during source refinement stage using the Ferret genetic algorithm. } \label{BLplot2}
\end{figure}

Ferret's convergence on the source refinement problem is shown in Figure \ref{BLplot2}. The smooth convergence curve is a hallmark of linear or other easy problems.  We have noted that the source reconstructions begin fitting to noise in the target image slowly, so it is generally quite easy to find an acceptable termination criteria for the algorithm. Since each pixel in the source is independently treated by the GA, this problem cannot be expressed in terms of the SVD expansion in the same way that the solutions to a linear optimization step can.  However, to quantitatively ensure that overfitting to noise is prevented, we once again form the L-curve between the image $\chi^2$ and a linear regularization measure $\sum s_i^2$ to quantify the amount of noise in the source.  In practice the L-curve analysis in the final analysis step is of limited use due to the smooth convergence of the algorithm and the slow rise in noise in the reconstructed sources.  Ferret is able to converge to a source near the location of the true solution for all situations that we have tested.  The best solution typically agrees with the true source to within $15\%$ though we have noticed variation in the details of the derived sources from run to run, which is expected considering the large number of parameters involved in the optimization.  It is interesting, and perhaps surprising, that the Locust PSO is unable to solve this problem, despite its linearity. We conclude that a GA is a more robust and efficient approach than particle swarm optimization for both of these optimization problems.  When the problem is small, a PSO can often find the solution in a comparative amount of time as a GA.

\section{Conclusions}
\label{sec:conclusions}
The semilinear method provides an elegant way to describe gravitational lens inversion in terms of a least squares problem, but is limited to relatively small images and a narrow PSF. This is due to the fact that the semilinear method requires the inversion of a large matrix whose size increases as the fourth power of the number of source pixels, and the sparsity of this matrix is reduced as larger PSFs are used. Solving for lens parameters is a nonlinear optimization problem, which can be solved by global optimization techniques. We applied and compared the Ferret GA and Locust PSO to determine the nonlinear parameters of the lens model.  The global optimization of lens parameters requires a lens inversion for each set of lens parameters tested, and $10^4-10^5$ such evaluations are required for a thorough exploration of the parameter space and mapping of the optimal region. This reinforces the need for fast lens inversion techniques that scale well with the size of the image and PSF.

We addressed the need for a fast lens inversion algorithm by developing a matrix-free approach to solve the least squares lensing problem, based partly on recent developments in the image deblurring literature, which solves the problem without the need to explicitly build the lens or blurring operators. This approach is intended to complement the semilinear method when speed is of the essence, or when large images and broad, highly structured PSFs are used. We note that our approach can be extended to the case of a spatially variant PSF.  Our analysis evaluated the convergence behavior of a matrix-free method using several local optimization methods. We found that the CGLS method is fastest to converge, but all linear optimization schemes suffer from over-fitting of noise if the optimization is not stopped at the critical iteration, which cannot be predicted a priori. We showed that steepest descent methods are more robust against over-fitting to noise at the expense of the speed of convergence.

The number of degrees of freedom in the iterative optimization step is estimated using a Monte Carlo method, allowing us to draw connections to the work of \citet{suyu06} that estimate the number of degrees of freedom using Bayesian statistics. We derived a formula for the number of degrees of freedom based on the filter factors of the Tikhonov regularization problem, which agrees with the expression found by \citet{suyu06} using Bayesian analysis.

We developed a novel method that computes the optimally regularized solution for each set of lens parameters by finding the point of maximum curvature in the trade-off curve between $\chi^2$ and a measure of the amount of regularization in the solution, which we took to be the sum of the squares of source pixel intensities.  The ambiguity of choosing a regularization parameter or stopping criteria is removed, because we automatically determine the optimal number of iterations (regularization constant) using the L-curve.  We evaluate the fitness of lens parameter sets using the image $\chi^2$ statistic.

The convergence and parameter space mapping properties of the Ferret GA and the Locust PSO schemes were compared, and we determined that the GA explores the parameter space more thoroughly than the PSO. The GA obtained a more detailed optimal set of solutions, highlighting the degeneracy in the position angle of a Singular Isothermal Elliptical lens model due to the rotational symmetry of the lens.  Both methods converge at a similar rate.

As a final refinement step in the image reconstruction our approach uses the GA or PSO to directly solve for pixel intensities.  This addition has the important benefit that the non-negativity of the source intensity profile can be enforced.  It is notable that the Ferret GA was able to solve this bounded linear solution refinement problem, but the Locust PSO failed due to the high dimensionality of the search ($\sim$2500 parameters). This analysis step shows stable convergence, and noise is introduced to the source very slowly. In practice this routine is relatively insensitive to stopping criteria.

This paper serves as a foundation for future explorations, which will apply the techniques discussed here to data, and expand them to include non-parametric lens models, such as those discussed by \citet{VK09} and \citet{Saha}. Non-parametric lens density models are extremely valuable, since dark matter haloes may contain significant substructure \citep{K05} that is not taken into account by analytical lens models. GAs have been applied to this problem previously, specifically by \citet{Lies07}, using the work of \citet{Diego} as a starting point. \citet{Lies09} used such an approach to model the system SDSS $J1004+4112$. This approach could be used in conjunction with the semilinear method to model complicated non-parametric lens density distributions and reveal the details of lensed extended sources.

\section{Acknowledgements}
A.R. acknowledges Bill Kocay for encouraging the exploration of matrix-free methods in the context of this work, Maiko Langelaar for technical support, and NSERC for funding this research. J.F. acknowledges funding from an NSERC Discovery Grant.  A.R. and J.F. thank the anonymous referee, whose thoughtful comments significantly strengthened our work.  The authors dedicate this work to the memory of J. J. Guy Durocher.

\appendix
\section{Ferret and Locust Details}

This Appendix offers a few additional details about the Ferret and Locust optimizers used in this paper and how their strategy parameters were set.  A more complete description is given in the Qubist User's Guide \citep{Fiege}, which is available from J. Fiege.

\subsection{Ferret Genetic Algorithm}
Most GAs encode model parameters on binary strings \citep{holland75, goldberg89}, with mutations and crossovers defined as operators that work directly on these strings.  For example, a mutation would typically flip a single bit, while a simple crossover would cut two binary strings at the same position and exchange the parts of the string to the right of the cut, effectively mixing together two individuals in the population.  If these strings represent real valued parameters of a model, it is necessary to decode the binary representation into real numbers prior to evaluation.  Ferret is specialized to work directly with genotypes specified by a list of real-valued parameters, thus side-stepping the conversion from binary strings to real numbers.  An individual in Ferret is therefore represented by a point in an $N$-dimensional real vector space, where $N$ is the number of parameters or ``genes'', which allows more elaborate mutation and crossover operators than can be defined on a simple binary string.

Ferret contains many options, which are controlled by ``strategy parameters'' that are encoded in a MATLAB structure called {\tt par}.  The strategy parameters are defined by a setup file, which is read at the start of a run.  Ferret contains a default setup file, which fills in any strategy parameters not specified by the user.  These default values are often adequate and the software is not usually very sensitive to the exact choice of strategy parameters.  This robustness is achieved in part by an adaptive algorithm that automatically controls several of the most important control parameters, affective mutations and crossovers.

A Ferret run evolves {\tt par.general.NPop} populations, where the size of each population is set by {\tt par.general.popSize}.  Ferret uses a single population by default, and it is recommended to set the population size in the range of $100-500$.  Generally, this choice is guided by the computational expense of evaluating the fitness function, the complexity of the problem, and the user's experience solving it.  Larger populations tend to explore the parameter space more thoroughly than smaller populations, but at greater cost.  When {\tt par.general.NPop > 1}, the populations interact weakly with each other by exchanging individuals with probability {\tt par.immigration.PImmigrate $\approx$ 0.01} each generation.  This is beneficial for some very difficult problems because multiple populations explore the parameter space almost independently, thus increasing the probability of finding the global solution.  Ferret often performs better on very difficult problems when the total number of individuals is divided into several populations rather than placing them all into a single population.  However, we used a single population with {\tt par.general.popSize=200}.

Ferret's mutation operator is defined as a perturbation in an $N$-dimensional real vector space, where the magnitude of the perturbation is drawn from an initially Gaussian distribution, whose standard deviation is determined by a strategy parameter {\tt par.mutation.scale=0.25} by default.  The distribution of mutation scales is under adaptive control, and evolves during each run, as Ferret preferentially selects values that result in improved fitness.  Ferret's default mutation rate is given by {\tt par.mutation.PMutate=0.05}.

The role of crossover in a GA is to mix together two different solutions to produce offspring that are intermediate between the parents.  Ferret contains two different crossover operators, which mix genes in fundamentally different ways.  Ferret's ``X-type'' crossover operator is a geometry-based operator that can be shown to be analogous to the bit string operator found in traditional binary encoded GAs.  X-type crossover is essentially an averaging operation, which draws a line between the parameter space coordinates of two individuals and selects a point between the individuals on that line.  The fractional distance traveled along this line is drawn from a distribution, which was initialized to a Gaussian random distribution of standard deviation {\tt par.XOver.strength=0.25} at the beginning of the run.  The distribution of crossover strengths is under adaptive control and co-evolves with the population to prefer crossover strengths that tend to result in improved fitness.  Note that it is possible to occasionally overshoot during a crossover by drawing a crossover strength greater than one.  Surprisingly, this turns out to be beneficial on many problems because it helps to expand the population into long, slender valleys by occasionally overshooting the end points of the distribution.  X-type crossover is Ferret's primary search mechanism, so we normally set {\tt par.XOver.PXOver=1} to set the crossover probability to 100\%.

Ferret's ``building block crossover'' operator is at the heart of its linkage-learning system and has no analogy in traditional GAs.  This type of crossover exchanges a building block, or a group of parameters previously identified as linked, in their entirely from one individual to another.  Building block crossover efficiently propagates building blocks responsible for high-quality solutions throughout the population and mixes them with other high-performing building blocks comprised of other parameters.  We normally set {\tt par.XOverBB.PXOver=1}, which indicates a 100\% chance of mixing building blocks.

Ferret makes a duplicate copy of all populations prior to mutation and crossover, effectively doubling the number of individuals.  Ferret's selection operator is applied after the mutation and crossover operators, using a binary tournament scheme in which individuals are drawn randomly from the populations modified by mutation and crossover to compete against individuals drawn from the unmodified duplicate populations.  Individuals that win a tournament are allowed to propagate to the next generation and the losers are destroyed.  The probability of competition is normally 100\%, but it is possible to reduce the selection pressure by setting {\tt par.selection.pressure < 1}.  This delays convergence, thereby allowing more time for exploration, by causing Ferret to ignore fitness values during tournament selection with probability equal to {\tt 1-par.selection.pressure}.

Sometimes a second round of competition is required when individuals tie in a tournament.  This occurs commonly in multi-objective problems, when {\tt par.selection.pressure < 1}, or when a fuzzy tolerance has been defined for a single-objective problem.  For example, we map out some region of the parameter space within $\Delta\chi^2$ ($dchi2$) of the minimum value by setting {\tt par.selection.FAbsTol=dchi2} to tell Ferret to ignore differences in fitness less than this amount.  In this case, Ferret employs a niching strategy similar to that discussed by \citet{fonseca93}, which prefers solutions with fewer near neighbors over solutions with a greater number of neighbors.  The logic behind this preference is simple: solutions in a less populated region of parameter space are more unique, and therefore more valuable to the exploration of the space.

\subsection{Locust Particle Swarm Optimizer}
Locust is a relatively simple code to configure, compared to the myriad of options allowed by Ferret.

The most important strategy parameter controlling a PSO is the number of particles in the swarm, given by {\tt par.swarm.N}.  In general, larger swarms tend to explore the parameter space more thoroughly, but may require more time to do so.  Very small swarms are problematic because they may sample the space poorly and miss the global solution.  There is no established rule for choosing the swarm size.  One typically starts with about 100 particles and decreases the number of particles if experience shows that this decreases the run time without causing problems with reliability.  Very difficult problems may require more than 100 particles, and we used {\tt par.swarm.N=200}.

{\tt par.swarm.cg} and {\tt par.swarm.cp} are, respectively, the global best and personal best constants used in Equation (\ref{eq:PSO}).  Both of these parameters should be of order unity, but setting $cg$ slightly less than $cp$ is usually helpful because this places more emphasis on exploration of the parameter space because the particles are influenced less by the global best solution.  Increasing $cg$ relative to $cp$ places more emphasis on the exploitation of the global solution or solutions, at the expense of exploration, because all particles will be drawn to the optimal region more rapidly.  We used the default values: {\tt par.swarm.cg=0.5} and {\tt par.swarm.cp=1}.

{\tt par.swarm.dt} is the time step between updates to the swarm positions and velocities.  Therefore, the time step $dt$ affects the rate of sampling of the parameter space as particles move around on their orbits, but has no effect on the accuracy of the obits because Locust uses an exact solution to the simple harmonic oscillator orbit equations approximated by the finite difference equation given by (\ref{eq:PSO}).  We used the default value {\tt par.swarm.dt=1}.

PSOs require damping to cause the particle swarm to settle down to a converged solution.  Locust is designed such that {\tt par.swarm.TDamp=1} corresponds to a critically damped harmonic oscillator.  Generally, underdamped oscillations are required so that multiple orbits explore the parameter space before the swarm converges.  We used the default value for the damping time {\tt par.swarm.TDamp=10}.

\clearpage
\newpage
\begin{table}
\begin{center}
\begin{tabular}{ | l | l | l | l | l | l |}
\hline
 Best Fit Solution & Ferret GA & Locust PSO & True & Lower Limit & Upper Limit \\ \hline
 $\chi^2_r$ & 1.010 & 1.012 & 0.998 & - & - \\ \hline
 $\sigma_v$ & 260.002 & 259.892 & 260.000 & $250.000$ & $280.000$ \\ \hline
 $\epsilon$ & 0.401 & 0.399 & 0.400 & $0.200$ & $0.500$ \\ \hline
 $x$ & $-2.101\times10^{-4}$ & $-2.123\times10^{-4}$ & 0.000 & $-0.500$ & $0.500$ \\ \hline
 $y$ & 0.119 & 0.120 & 0.120 & $-0.500$ & $0.500$ \\ \hline
 $\theta_L$ & 4.713 & 4.713 & $\pi/2$ & $0.000$ & $2\pi$ \\
 \hline
\end{tabular}
\end{center}
\caption{Optimal lens parameters found by the Qubist optimizers for the example given in the text using the SIE lens. We restrict the range of the lens parameters to prevent convergence to a trivial solution. Performance of the GA and PSO routines is similar as can be seen from the reduced image $\chi^2$.}
\label{tab:FerretLocustComp}
\end{table}

\clearpage
\newpage

\clearpage
\newpage

\clearpage
\newpage

\clearpage
\newpage

\clearpage
\newpage

\clearpage
\newpage

\clearpage
\newpage

\clearpage
\newpage

\clearpage
\newpage

\clearpage
\newpage

\clearpage
\newpage

\clearpage
\newpage

\end{document}